\documentclass{article}

\usepackage[english]{babel}

\usepackage[letterpaper,top=2cm,bottom=2cm,left=3cm,right=3cm,marginparwidth=1.75cm]{geometry}

\usepackage{amsmath}
\usepackage{graphicx}
\usepackage[colorlinks=true, allcolors=blue]{hyperref}
\usepackage[auth-sc, affil-it]{authblk}
\usepackage{caption}
\usepackage{amssymb}
\usepackage{tabularx}
\usepackage{arydshln}

\usepackage{xr}
\usepackage[numbers,sort&compress]{natbib}

\usepackage[framemethod=TikZ]{mdframed}
\usepackage{lipsum}

\usepackage[normalem]{ulem}

\usepackage{parskip}
\mdfdefinestyle{MyFrame}{%
    linecolor=blue,
    outerlinewidth=1pt,
    roundcorner=2pt,
    innertopmargin=\baselineskip,
    innerbottommargin=\baselineskip,
    innerrightmargin=20pt,
    innerleftmargin=20pt,
    backgroundcolor=white!50!white}

\providecommand{\keywords}[1]
{
  \small	
  \textbf{\textit{Keywords---}} #1
}

\providecommand{\listofabbreviation}[1]
{
  \small	
  \textbf{\textit{List of abbreviations---}} #1
}

\title{Progress on Data-Driven, Multi-Objective Quantum Optimization}

\author[1,*]{Thomas Plehn}
\author[1]{Daniel Barragan-Yani}
\author[1]{Eric Breitbarth}
\author[1]{Guillermo Requena}
\author[1]{David Melching}

\affil[1]{Institute for Frontier Materials on Earth and in Space, German Aerospace Center (DLR), Linder Hoehe, Cologne, 51147, Germany}
\affil[*]{Corresponding author: Thomas.Plehn@dlr.de}

\begin{document}
\maketitle
 
\begin{abstract}
 
    Here, we present two complementary approaches that advance quadratic unconstrained binary optimization (QUBO) toward practical use in data-driven materials design and other real-valued black-box optimization tasks. 
    First, we introduce a simple yet powerful preprocessing scheme that, when applied to a machine-learned QUBO model, entirely removes system-level equality constraints by construction. This makes cumbersome soft-penalty terms obsolete, simplifies QUBO formulation, and substantially accelerates solution search. 
    Second, we develop a multi-objective optimization strategy inspired by Tchebycheff scalarization that is compatible with non-convex objective landscapes and outperforms existing QUBO-based Pareto front methods. We demonstrate the effectiveness of both approaches using a simplified model of a multi-phase aluminum alloy design problem, highlighting significant gains in efficiency and solution quality. Together, these methods broaden the applicability of QUBO-based optimization and provide practical tools for data-driven materials discovery and beyond.

\end{abstract}


\listofabbreviation{CGFM: Constraint-guided feature mapping, DDTS: Data-driven Tchebycheff scalarization, QUBO: Quadratic unconstraint binary optimization, FM: Factorization machine, QO: QUBO optimization, QA: Quantum annealing, QAOA: Quantum approximate optimization algorithm, SA: Simulated annealing, $\kappa$: Thermal conductivity, $E$: Young's modulus, $\rho$: density, $\alpha$: Linear coefficient of thermal expansion, $\Delta T$: Solidification interval}

\section{Introduction}
\label{sec_intro}

Solving many real-world problems critical for economic and scientific progress, such as discovering new innovative materials, finding optimal logistics strategies, planning the best travel routes, or designing the most efficient duty schedules, often involves tackling complex combinatorial optimization tasks. Quantum optimization is a promising route to deal with the exponential explosion of computational effort required to solve such problems. It does so by leveraging quantum mechanics (specifically quantum tunneling and superposition effects) to explore a vast search space more efficiently than classical methods \cite{Preskill2018,Yarkoni2022}. Quantum optimization essentially means solving a quadratic unconstrained binary optimization (QUBO) problem \cite{Lucas2014}. Although conceptually alternative approaches do exist, most recent research, focused on finding practical applications for quantum optimization, is built  on the QUBO-based methodologies: quantum approximate optimization algorithm (QAOA) \cite{Farhi2014, Zhou2020} and quantum annealing (QA) \cite{Johnson2011}. On the one hand, the D-Wave QA technology seems to be a very promising candidate for near to mid-future industrial scale application as it provides more than 5,000 physical qubits \cite{dwave_advantage_overview, Boothby2020, King2025}. On the other hand, circuit-based methods like QAOA provide extraordinary flexibility for algorithm engineering, which promises more rapid and continuous improvements. In addition to QAOA, the use of variable inverse time evolution has been repeatedly proposed as a promising means for solving combinatorial problems \cite{McArdle2019, Morris2024}.

In order to meet the mathematical QUBO  formulation, considered optimization problems must first be binary encoded. Unfortunately, many realistic systems involve higher-order interactions among variables, which cannot directly be mapped on a quadratic model. Turning such problems into QUBO form usually comes at the cost of restricting binary interactions to second order, or involves the introduction of additional auxiliary variables \cite{Anthony2017, Mandal2020}. Thankfully, data-driven, machine-learning-aided QUBO optimization ansatzes can treat realistic use-cases within the QUBO formalism efficiently \cite{Wilson2021}. Specifically, we highlight the potential of the data-driven FM+QO method for applied problems \cite{Kitai2020, Kim2022, Kim2023, Urushihara2023, Nawa2023}. 
It was developed to  combine the efficiency of either QA, QAOA or any QUBO-based quantum optimization algorithm, with the powerful quadratic regression of factorization machines (FM) \cite{Rendle2010}.
The core idea of FM+QO is the following: starting with an initial set of data points for the objective to be optimized, a FM surrogate regression model is trained to predict a target property. Then, the QUBO optimization (QO) method part optimizes the corresponding QUBO. The discovered optimum is checked against a validation source, e.g., by simulation or experiment, and finally added to the data set. In this way, the FM model is iteratively refined in an overarching active learning process. Various studies have shown the great potential of FM+QO with QA as a black-box optimizer in the field of materials research, for example, applied on optimization of optical multilayer systems and metamaterials \cite{Kitai2020, Kim2022, Kim2023}, core-shell nanoparticles \cite{Urushihara2023} or the functional optimization of atomistic systems \cite{Nawa2023}.


Despite all efforts, however, a practical advantage of quantum-computing-based QUBO solvers against state-of-the-art classical optimizers could not be demonstrated yet, particularly owing to thermal noise induced inaccuracies and other imperfections feeding decoherence within gate-based quantum hardware or QA devices \cite{Yarkoni2022, Arai2023}. While hardware developments will undoubtedly continue to advance and improve solution quality, it is the co-development of algorithms that must extract the full potential of today's quantum devices.


In this regard, especially the numerical introduction of necessary constraints describes an Achilles heel of many current QUBO formulations for quantum solvers and is recently a hot topic \cite{Hen2016, Ohzeki2020, Yu2021, Mirkarimi2024, Binninger2024, Montanez2024}. Although constraints appear in many real-world optimization application cases, they cannot be intrinsically embedded in the QUBO solution. 
An ubiquitous type of constraint is the one which enforces a constant sum of the input variables. 
An example is the optimization of material compositions as a mixture of different base constituents. Naturally, the solutions must satisfy the systemic equality constraint that the mixing ratios, $r_i$ [\%], of $N_\text{comp}$ (input) 
base components always sum properly to $\sum_i^{N_\text{comp}} r_i = 100.0~\%$.
Classically, constraints are treated by perturbing the problem-specific QUBO with a quadratic regularization often referred to as \textit{soft constraint}.
However, such soft constraints can make the resulting QUBO problem substantially harder to solve with QA or QAOA, owing to a narrowing of the minimum energy gap that isolates the aspired problem adiabatic ground state solution from other solutions \cite{Miessen2024, Zurek2005, Hen2016, Roch2023}.
Beyond, soft constraints pose particular challenges on current QA devices due to two well-known technical limitations \cite{dwave_advantage_overview, dwave_ocean_docs, Montanez2024}. First, any additional interaction term inflates the minor embedding needed to map the QUBO to the hardware architecture graph. And second, the analog setting of Ising spin biases and couplings operates within a limited dynamic range. In order to ensure feasibility, soft constraints usually tend to dominate this range and in consequence suppress the effective precision of the remaining terms. In practice, each constraint requires careful tuning to balance its intended effect against its detrimental impact on solution quality, however, this typically relies on heuristic trial-and-error processing.
More improved alternative ways to incorporate constraints range from proposals of grand-canonical linear correction of optimization cost function \cite{Binninger2024, Binninger2025}, that can greatly reduce the required strength of quadratic penalties, to ansatzes with purely linear penalties \cite{Ohzeki2020, Yu2021, Mirkarimi2024} easing embedding demand but at the same time making parameter tuning essentially more difficult and susceptible to the often noisy and complex optimization landscape. A fundamentally different theoretical ansatz has been developed for adiabatic state propagation \cite{Hen2016, Hadfield2019}. By choosing the right initial Hamiltonian for the problem-specific QUBO at hand it avoids constraints altogether. This comfort, however, is paid by a non-trivial state initialization \cite{Wang2020,Niroula2022} and an expensive mixer unitary which is impractical for present QA devices.


In this paper, we introduce two methods to advance data-driven quantum optimization. The first method, which we call \textit{constraint-guided feature mapping} (CGFM), is in the spirit of Hen et al.~\cite{Hen2016}. It describes a novel pre-processing scheme for avoiding explicit treatment of equality constraints when solving real-valued, data-driven optimization problems with FM+QO. To do this, we exploit the feature of the data-driven FM+QO optimization method of not relying on a fixed known analytical QUBO model expression. Briefly, the method makes constraints unnecessary, because QUBO modeling and QO exclusively takes place within the sub space of feasible solutions. In addition, since the CGFM is part of our pre-processing, it remarkably retains its effectiveness even in the case of using quantum solvers on NISQ hardware. Although real-valued optimization problems pose a significant challenge for current QUBO-based quantum solvers due to their need to discretize continuous variables, we expect that addressing such problems becomes an increasingly important focus as quantum computing progresses beyond the NISQ era \cite{Arai2023}.

Our second method, which we refer to as \textit{data-driven Tchebycheff scalarization} (DDTS), addresses another blind spot of current QUBO optimization regarding real-world applications: the handling of multi-objective Pareto optimization. The search for efficient algorithms to find Pareto fronts remains an urgent and open area of research in many fields. Solving multi-objective optimization problems is more complex than single-objective tasks because the properties to be optimized usually conflict with each other \cite{Emmerich2018}. Instead of a single solution, one therefore searches for the Pareto front - a set of non-dominated solutions. As a state-of-the-art classical method, metaheuristic evolutionary algorithms provide efficient solution sets in short time \cite{Deb2002, Emmerich2018, Olvera2023}. However, with increasing problem size, metaheuristics suffer from longer computation times and lower solution quality. In addition, such population-based algorithms stumble if the many required objective function calls cannot be executed efficiently. In many industrial fields this is exactly the case. Compared to classical metaheuristics, quantum optimization can, in principle, handle exponential scaling of search spaces and, in particular, QA provides solutions within milliseconds \cite{dwave_ocean_docs}. 

With the advent of mature quantum hardware, advantage through quantum optimization will not stop at multi-objective problems. Therefore, concepts for quantum Pareto optimization algorithms \cite{Schworm2024, Aguilera2024, Ekstrom2025, Kotil2025}, in particular multi-objective QUBO ansatzes \cite{Imanaka2021, Schworm2024, Aguilera2024, Kotil2025}, are of recent interest. In principle, the way to describe a multi-objective optimization in the QUBO formalism by means of scalarization is well known from literature \cite{Pereira2022, Ikeda2024}. In such proposals, one attributes multiple QUBO models with a normalized preference vector and merges the weighted expressions into a single QUBO. This state-of-the-art "weighted sum" method, however, has a practical pitfall that can make its implementation rather tedious. Namely, setting the preference weights for scalarization is extremely case-specific. Moreover, we identify severe problems using the FM+QO method to sample solutions from the entire Pareto front, which we associate with the generally well-known weakness of the weighted-sum ansatz in non-convex situations.

Our DDTS ansatz circumvents the QUBO scalarization by making use of the Tchebycheff technique, which maps the multi-objective space vectors to a single objective. The Tchebycheff technique is well known from Pareto optimization problems of analytical type, where it properly complies also with non-convex functional expressions \cite{Pardalos2017}. Importantly, when starting from analytical QUBO expressions it would actually not be usable for scalarization owing to its specific operational form. Here, however, we again leverage the fact that the QUBO model is learned from data and apply the Tchebycheff technique directly on the data set preliminary to the FM training. 
To the best of our knowledge, neither the FM+QO method specifically nor the Tchebycheff scalarization in general have been used yet to solve QUBO-based multi-objective optimization tasks. Our demonstration of a data-driven QUBO-based Pareto optimization is therefore the first of its kind.

To demonstrate the strength and potential of the CGFM and DDTS ansatzes, we introduce a real-world materials science optimization problem.
Our new methods stand out particularly in the optimization of material microstructures for three reasons. First, such optimizations have in common the requirement for a systematic constraint deciding about design feasibility.
Second, they are typically multi-objective because they can show counteracting target properties. And third, in this field non-convex optimization problems are ubiquitous. In fact, thermal and mechanical material properties are ultimately based on energetic landscapes, which can be complex structured with local minima, energetic barriers and extended plateaus resulting in non-convex dependencies of system variables.

For decades, aluminum alloys have been the focus of extensive research due to their combination of low density, mechanical strength, and corrosion resistance, leading to applications in numerous industrial fields \cite{Hirsch2013,Polmear1999}. Aluminum alloys are metallic materials in which aluminum is the base matrix and principal constituent (typically 80-90 \%) accompanied by the addition of other elements such as  Mg, Si, Cu, Zn, Mn, Ni, Fe, depending on the required properties profile.
Depending on the composition and processing conditions, primary and/or secondary (intermetallic) phases can form and their three-dimensional morphology influences the properties of the alloys \cite{Bugelnig2024}. While understanding and predicting the formation of the microstructure during processing of Al alloys is a well-established field of research, the opposite approach, namely designing realistic microstructures that fulfill target properties and then infer the composition and processing path of the alloy is a promising reverse-engineering approach that has received less attention \cite{Khatamsaz2026}. However, the design of realistic microstructures for engineering alloys is a complex multi-scale issue that requires considering microstructure descriptors from the atomistic- (e.g. dislocations, vacancies, solutes, crystallographic orientation) through nanometer- (e.g. precipitates) to the micro-scale (e.g. grains, primary intermetallic phases). 

As a proof-of-concept and a first building block for this reverse-engineering approach for materials design, we consider in this work the microstructural optimization of a hypothetical Al alloy with five microstructural phases at the µm level, that can be found in cast Al-Si alloys: Al-matrix, eutectic Si, Mg$_2$Si, Al$_3$Ni and Al$_2$Cu. As target properties of the alloys we choose five material thermal and mechanical properties: thermal conductivity ($\kappa$), Young's modulus ($E$), density ($\rho$), linear coefficient of thermal expansion ($\alpha$) and the solidification interval ($\Delta T$). 
To limit computational cost and reduce model complexity, we employ rule-of-mixtures estimates to obtain quantitative, rapid predictions of $\kappa$, $E$, $\rho$, and $\alpha$. For $\Delta T$, we introduce a coarse approximation derived from the binary Al–Si phase diagram. These five simplified property models were then used to generate synthetic data of randomly sampled alloy designs and their associated properties, providing initial input for our data-driven FM+QO framework.

\section{Results and Discussion}

\subsection{Methodological Developments}

\begin{figure}[t]
    \centering
    \includegraphics[width=\textwidth]{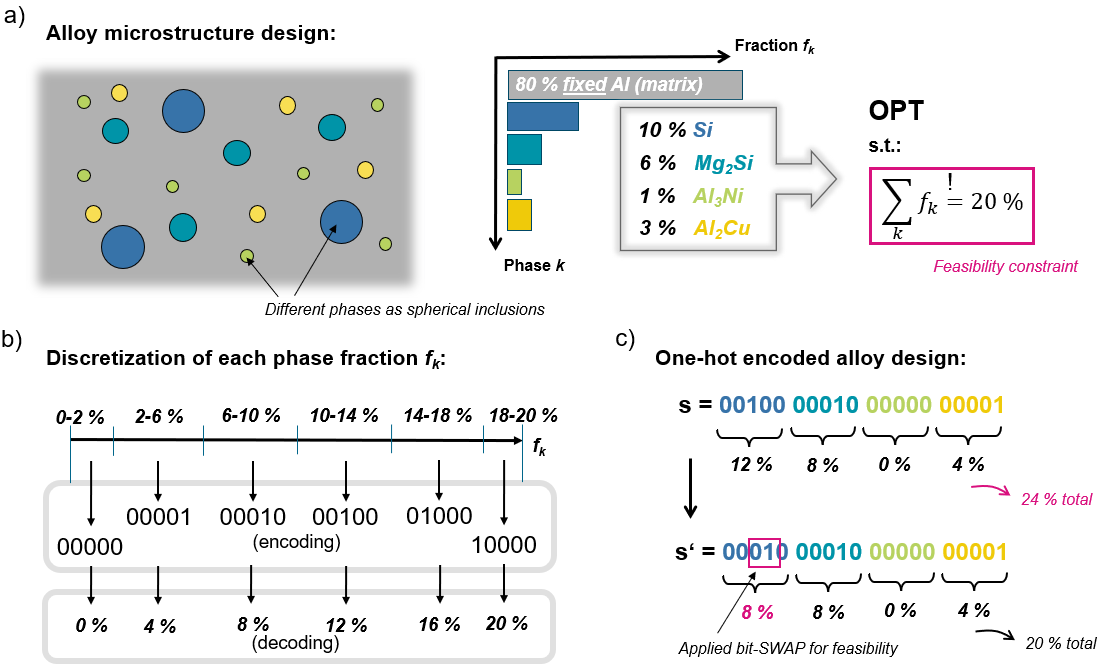}
    \caption{Illustration of the multi-phase aluminum alloy model and its encoding. a) Scheme of the simplified microstructure and phase morphology assumed for straight forward prediction of different thermal and mechanical material properties with the rule-of-mixtures: The Al-matrix dominate has the largest fraction with additional minor phases embedded in it. These phases are treated as isotropic and homogeneously distributed spherical particles (Si, Mg$_2$Si, Al$_3$Ni and Al$_2$Cu). The matrix fraction is fixed to 80 \%, while the four minor fractions, $f_k \in [0~\%,20~\%]$, are optimization targets under the given constraint of feasible alloy designs. b) Discretization of the continuous $f_k$ to prepare the binary encoding (top). Each increment is encoded uniquely using the one-hot encoding (middle). Each one-hot code is uniquely decoded into continuous $f_k$ again (bottom). Due to the finite encoding fidelity, the decoded fractions have limited accuracy. c) Translation from fractions to the alloy design encoding with the microstructure configuration in figure a.
    As indicated, the limited accuracy results here in an infeasible solution once decoded back to continuous $f_k$. Those special cases are identified and treated by a curing bit-SWAP applied to a randomly chosen $f_k$. (Please note, this correction is only required on the randomly generated alloy designs in the initial data set. The optimization operates in the encoded search space.)}
    \label{fig:discretization}
\end{figure}

Our multi-phase alloy model is schematically illustrated in Fig.~\ref{fig:discretization}. In a nutshell, it is prepared by first introducing continuous variable fractions, $f_k$ with $k=1,...,N_{\rm ph}$, for the $N_{\rm ph}=4$ secondary phases: eutectic Si, Mg$_2$Si, Al$_3$Ni and Al$_2$Cu. In our model we define a fixed Al matrix content of 0.8, i.e., $f_k\in \small[0,0.2\small]$. For convenience, we continue with normalized phase fractions, which are finally one-hot encoded to binary variables. A discretization of the continuous variables is vital for this step. We elaborate on these processes in Sec.~\ref{sec:encoding}.

\subsubsection{Constraint-guided feature mapping (CGFM)}
\label{sec_CGFM}

Let $\bar f_k$ denote the \textit{normalized} continuous variable phase fractions for the Al-alloy with $\bar f_k \in [0,1]$ and $k=1, \dots, N_{\rm ph}$.  
Obviously, the fractions must sum to 100\%, i.e.,
\begin{align}
\sum_{k=1}^{N_\text{ph}} \bar f_k = 1,
\label{sum_condition}
\end{align}

Any deviation from that would render an infeasible alloy composition. In order to incorporate this requirement in a QUBO optimization, one typically introduces soft equality constraint terms similar to
\begin{align}
C(\vec{x}) = \Big( \sum_{k=1}^{N_\text{ph}} \sum_{i=1}^{N_\text{bits}} \alpha_i x_i^{(k)} -  1 \Big)^2,
\label{eq_constraint}
\end{align}
already based on binary variables (or bits), $x_i^{(k)} \in \{0,1\}$, used to encode the $\bar f_k$ in Eq.~\ref{sum_condition} and the corresponding encoding-specific weights $\alpha_i$ (see Eq.~\ref{encoding} in Sec.~\ref{sec_enc_details}). The first sum runs through all $N_\text{ph}$ phases and the second one sums over all $N_\text{bits}$ binary variables used to encode one fraction variable.

In general, additional constraint terms in QUBO models lead to a more rugged structure of the potential energy surface of the system which, provided with many local minima, increases the probability that the optimization  gets stuck in a local minimum instead of finding the global one  \cite{Arai2023}. Furthermore, soft constraints increase configuration effort, as they have to be added with a strength factor, $\lambda$, which must be precisely tuned against the original QUBO model, and apparently this is exacerbated when several constraints have to be taken into account. 
These two disadvantages generally occur with quantum QUBO solvers and classical heuristics, but equality constraints of type Eq.~\ref{eq_constraint} are especially critical and detrimental to the solution quality of quantum algorithms (e.g., QA \cite{Binninger2024}).

In fact, the individual coupling terms in Eq.~\ref{eq_constraint} range over many orders of magnitude with realistic model scaling. When applying the one-hot type of encoding with $\alpha_i = i / N_\text{bits}$, this can be observed straightly from 

\begin{align}
\max \small\{ \alpha_i \small\} / \min \small\{ \alpha_i \small\} &= N_\text{bits}.
\label{min_max_ratio}
\end{align}
In fact $N_\text{bits}=101$ are required with one-hot encoding in order to model our phase fractions, $\bar f_k$, with less than 0.01 resolution. According to the ratio, Eq.~\ref{min_max_ratio}, the coupling terms in Eq.~\ref{eq_constraint} would range over 4 orders of magnitude after squaring.

As described in Sec.~\ref{sec_intro}, such a wide range of coupling terms is technically problematic regarding the analog nature of current QA devices. Because the strength factor of the constraint, $\lambda$, must be sufficiently large to guarantee feasibility, these terms often dominate the original QUBO contributions. When ranging over several orders of magnitude, the remaining QUBO coefficients become very small and increasingly prone to hardware noise. Moreover, the quadratic structure of Eq.~\ref{eq_constraint} couples all binary variables across all fractions. The resulting fully connected constraint graph is the worst case scenario for QA hardware, which makes ultimately long qubit chains for minor embedding of the QUBO unavoidable. These break more often and degrade solution quality despite post-processing. Gate-model platforms face a similar challenge. There, compilers insert SWAP operations to mediate interactions not supported by the hardware topology, increasing circuit depth and susceptibility to noise.

\begin{figure}[t]
    \centering
    \includegraphics[width=0.75\textwidth]{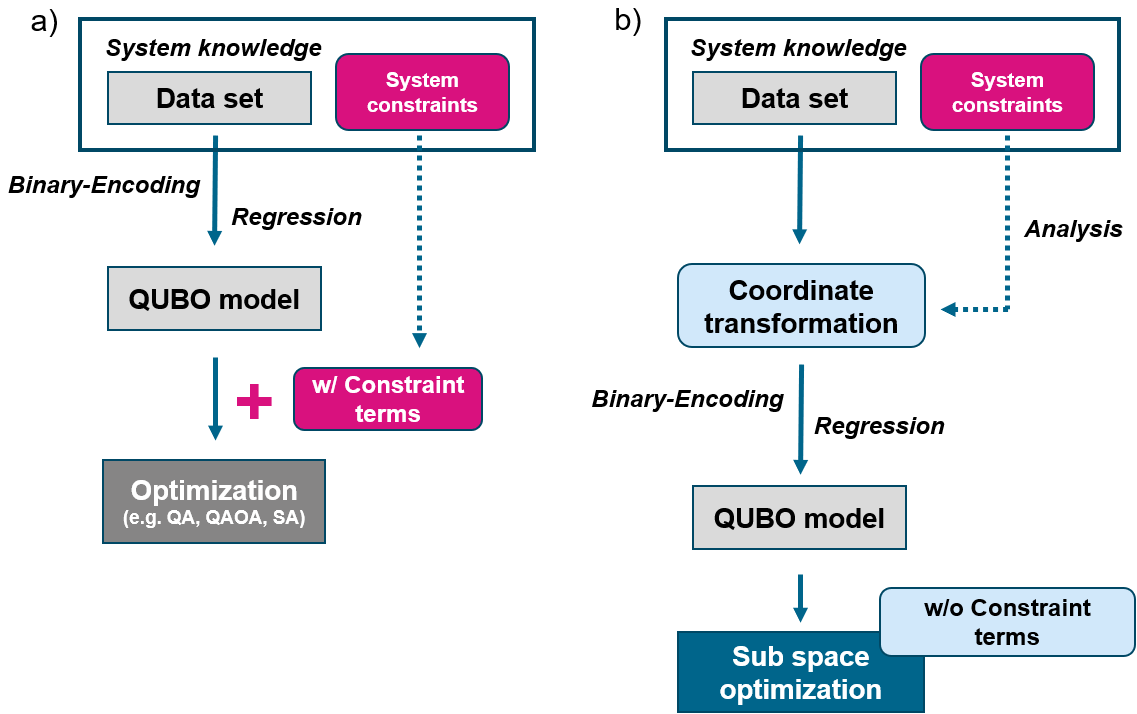}
    \caption{Comparison of the CGFM ansatz against current data-driven QUBO modeling. a) Current state-of-the-art: A binary quadratic regression model (e.g., FM) is trained based on a data set and introducing a binary encoding for continuous system variables. Then, the regression model can be translated directly into a QUBO problem. The constraint terms are finally formulated according to the given feasibility conditions (e.g., Eq.~\ref{eq_constraint}), configured with suitable scaling and added to the QUBO problem. b) CGFM ansatz: By analysing the constraints a coordinate transformation is established which maps the original variable space (i.e., data set features) to the sub space of feasible system states. Restricted to this sub space the QUBO modeling and subsequent optimization can be executed without the constraints.}
    \label{fig:CGFM_scheme}
\end{figure}

Our constraint-guided feature mapping (CGFM) ansatz opens the possibility to avoid the constraints in Eq.~\ref{eq_constraint} altogether when a QUBO model is learned from data, e.g., by facilitating a factorization machine (FM).
The ansatz is schematically illustrated in Fig.~\ref{fig:CGFM_scheme}. The core idea is that differentiable constraints which satisfy regularity conditions define smooth submanifolds, which can be locally described using a reduced set of free parameters according to the \textit{implicit function theorem}. The constraints can thus be explicitly incorporated into the data set through prior feature mapping.
Feature mappings are a well-known pre-processing instrument from machine learning. Our approach is inspired by the kernel trick in support vector machines, where mapping the data into a higher-dimensional feature space enables more effective classification. This idea can be understood as a coordinate transformation of the system variables. However, we transform our original features, the phase fractions $\bar f_k$, to finally dismiss one dimension.
The transformed coordinates are then binary encoded (analogously to the current procedure in Fig.~\ref{fig:CGFM_scheme}a) and utilized for the QUBO training. By this dimensionality reduction, the surrogate QUBO model and the optimization are restricted to the subspace of feasible solutions making the system-related constraints obsolete. In general, this should simplify the search for global optima.
Furthermore, due to the reduced dimensionality of the search space, its encoding requires less binary variables, which should further accelerate the optimization process.

For a concrete example, we focus on the constraint in Eq.~\ref{sum_condition}. First, we realize that Eq.~\ref{sum_condition} defines a ($N_\text{elem}-1$)-dimensional submanifold in the $N_\text{elem}$-dimensional search space. Using the fact that the fractions $\bar f_k$ are non-negative, we substitute $\bar f_k=a_k^2$ to transform Eq.~\ref{sum_condition} to the relation of an $N_\text{elem}$-dimensional sphere with radius $1$, i.e., $\sum_k a_k^2 = 1$. The CGFM ansatz now introduces a mapping which transforms the optimization problem from the original fraction variables $\bar f_k$ to angular coordinates $\omega_k$ parametrizing the sphere. The spherical coordinates are well-known and we refer to the Supplementary Information \ref{sec_CD_SO}  for further computational details.
Finally, by expressing the problem in the new coordinates $\omega_k$, the systemic constraint is satisfied by construction, ensuring that all downstream solutions remain feasible and eliminating the need to include additional constraint terms in the QUBO model.
Consequently, the complexity of handling the constraints during the optimization is now shifted to a potentially more complicated process of training the FM surrogate model. We discuss this further in Sec.~\ref{sec_so_opt}.

\subsubsection{Data-driven Tchebycheff scalarization (DDTS)}

\label{sec_DDTS}

Concerning QO, the state-of-the-art multi-objective approach to search for Pareto optimal solutions is to implement the \textit{weighted-sum} scalarization method (see Eq.~\ref{eq_QUBO} in Sec.~\ref{sec_fmqo_al}). The entire Pareto front would then result from enough optimal solutions of the scalarized objective, $g_{\vec{w}}(\vec{x})$ in Eq.~\ref{eq_QUBO}, with different weight vectors, $\vec{w}$, expressing the preference of the considered objectives.
For non-convex Pareto fronts, however, the weighted-sum method is unable to resolve the entire Pareto front \cite{Pardalos2017, Pereira2022, Ikeda2024}. The same challenge occurs when scalarizations of several objective-specific QUBO models produce non-convex Pareto fronts, thereby limiting the effectiveness of traditional weighted-sum approaches.

Although well-established alternative scalarization approaches exist for non-convex, continuous multi-objective optimization \cite{Pardalos2017}, those usually lead to QUBO-incompatible expressions or numerical procedures.
Recently, an alternative scalarization ansatz specialized for QUBO optimization has been proposed \cite{Ikeda2024}. The method is based on the principles of the penalty-based boundary intersection method and works on non-convex Pareto fronts. However, the QUBO expression to be solved contains product terms of the involved objective functions. So, to finally stay in QUBO format, the original objective functions are restricted to linear dependence on the binary variables, which is an apparent drawback regarding many relevant use cases.

In our data-driven Tchebycheff scalarization (DDTS) framework, in the same spirit as for CGFM, we implement the scalarization step to the data set in terms of a pre-processing and, therefore, before the mathematical QUBO form of the system model restricts further algorithm engineering. Consequently, the DDTS ansatz is not restricted to linear descriptions of our system. Furthermore, since it originates from the Tchebycheff scalarization technique, the approach works well for non-convex Pareto fronts. 

The (weighted) Tchebycheff scalarization \cite{Pardalos2017} is well-known for continuous variable multi-objective problems. In addition to preference  weights, $\vec{w} \in \mathbb{R}^m$, it involves a reference point in the objective space, $\vec{u} \in \mathbb{R}^m$. This so-called \textit{utopian point}, needs to fulfill $u_p < \min_{x \in \mathbb{A}} f^{( p)}(\vec{x})$, where $f^{( p)}(\vec{x})$ denotes the $p$-th objective function and $\mathbb{A}$ the (continuous variable) search space. The multi-objective Tchebycheff problem is then of the form
\begin{align}
\min_{x \in \mathbb{A}} \Big( \big[ \max_{p=1,...,m} w_p ( f^{(p)}(\vec{x}) - u_p)  \big] + \gamma \sum_{p=1}^m (f^{(p)}(\vec{x}) - u_p) \Big),
\label{eq_Tchebycheff_orig}
\end{align}
attributed with different weights, $\vec{w}$ (with $0<w_p <1$ and $\sum_p w_p = 1$). The entire Pareto front can be found for convex and non-convex problems. The second term in Eq.~\ref{eq_Tchebycheff_orig} quantifies the Manhattan distance of point $f^{(p)}(\vec{x})$ from the utopian in objective space. It regularizes the original Tchebycheff formulation to obtain only strong Pareto optimal solutions \cite{Pardalos2017, Steuer1983}.

\begin{figure}[t]
    \centering
    \includegraphics[width=\textwidth]{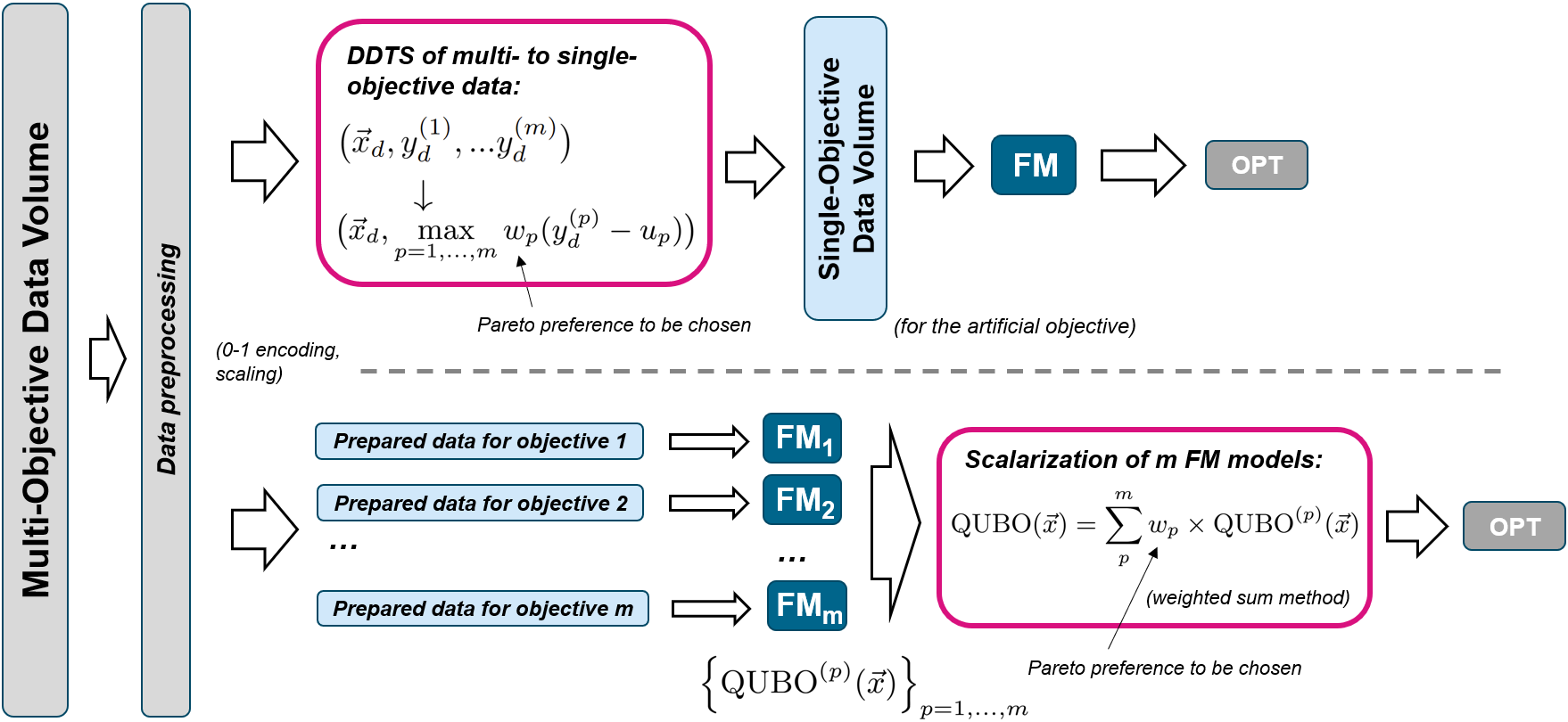}
    \caption{Schematic processing of our data-driven Tchebycheff scalarization (DDTS) method (upper part). The sequences are part of the FM+QO method and repeated for each iteration of the active learning. In the lower part, the corresponding workflow is given based on the current state-of-the-art weighted sum method for QUBO scalarization.}
    \label{fig_DDTS}
\end{figure}

Our multi-objective DDTS is motivated by the problem form of Eq.~\ref{eq_Tchebycheff_orig}. Fig.~\ref{fig_DDTS} illustrates the workflow. Since Eq.~\ref{eq_Tchebycheff_orig} is clearly not a QUBO problem, this scalarization technique cannot be utilized on our problems. However, we can circumvent this issue thanks to the data-driven ansatz and introduce the scalarization directly on the data, i.e., the multi-objective data set is translated into a data set with only a single \textit{artificial} objective (see upper part in Fig.~\ref{fig_DDTS}). After that, we train a single QUBO model problem on the data set with the artificial objective, instead of applying the weighted-sum approach on individual QUBOs trained each for a different objective function (see lower part in Fig.~\ref{fig_DDTS}).

Let $\vec{x}_d$ be the search space vector of the $d$-th data point, and $y^{(p)}_d$ the corresponding value of objective $p$, the DDTS ansatz for pre-processing of the data set follows as replacement according to
\begin{align}
(\vec{x}_d,y^{(1)}_d, ...y^{(m)}_d) \longrightarrow  (\vec{x}_d, \hat{y}_d).
\label{eq_DDTS_1}
\end{align}
By this procedure the data set becomes single-objective where $\hat{y}_d$ is defined as
\begin{align}
\hat{y}_d = \max_{p=1,...,m} w_p (y^{(p)}_d - u_p).
\label{eq_DDTS_2}
\end{align}
The DDTS is repeated within each active learning iteration. 
The numerical procedure works best after normalizing the range of objective values $y^{(p)}_d$. The utopian values are either known or can be chosen automatically based on knowledge from the data set. More details on the implementation are given in the Supplementary Information \ref{sec_CD_MO}.
Finally, the FM is trained on the data set $(\vec{x}_d, \hat{y}_d)$.
In extreme cases of $w_p=1$ and $w_{q\neq p}=0$, the artificial objective value $\hat{y}_d$ resembles $y^{(p)}_d$ shifted by $u_p$, i.e., the QUBO model is then trained to predict the objective values $y^{(p)}$ and solving the multi-objective QUBO resembles the single-objective optimization. In all other cases, the artificial objective of point $d$, $\hat{y}_d$, takes the value of the objective that is most distanced from its corresponding utopian point. This decision is made independently for each data point and the concrete weighting, $\vec{w}$, balances this preference.

Overall, the closer a data point $(\vec{x}_d,y^{(1)}_d, ..., y^{(m)}_d)$ lies to the utopian value along objective $p$, the higher the weight $w_p$ must be chosen in order to make that data point part of the FM training regarding the prediction of $f^{(p)}$. The number of points in the training data set which are attributed to objective $y^{(p)}$ decide how much the trained QUBO model is characterized by objective $f^{(p)}$. This effect allows DDTS to work as a multi-objective add-on for the FM+QO method. If one wishes to target a specific point on the Pareto front, the weights $\vec{w}$ can be adjusted, respectively. However, if the entire Pareto front is of interest, the preference weights can be chosen randomly for each iteration.

\subsection{Simulation results}

\subsubsection{Single-objective optimization with CGFM ansatz}
\label{sec_so_opt}

\begin{figure}[t]
    \centering
    \includegraphics[width=\textwidth]{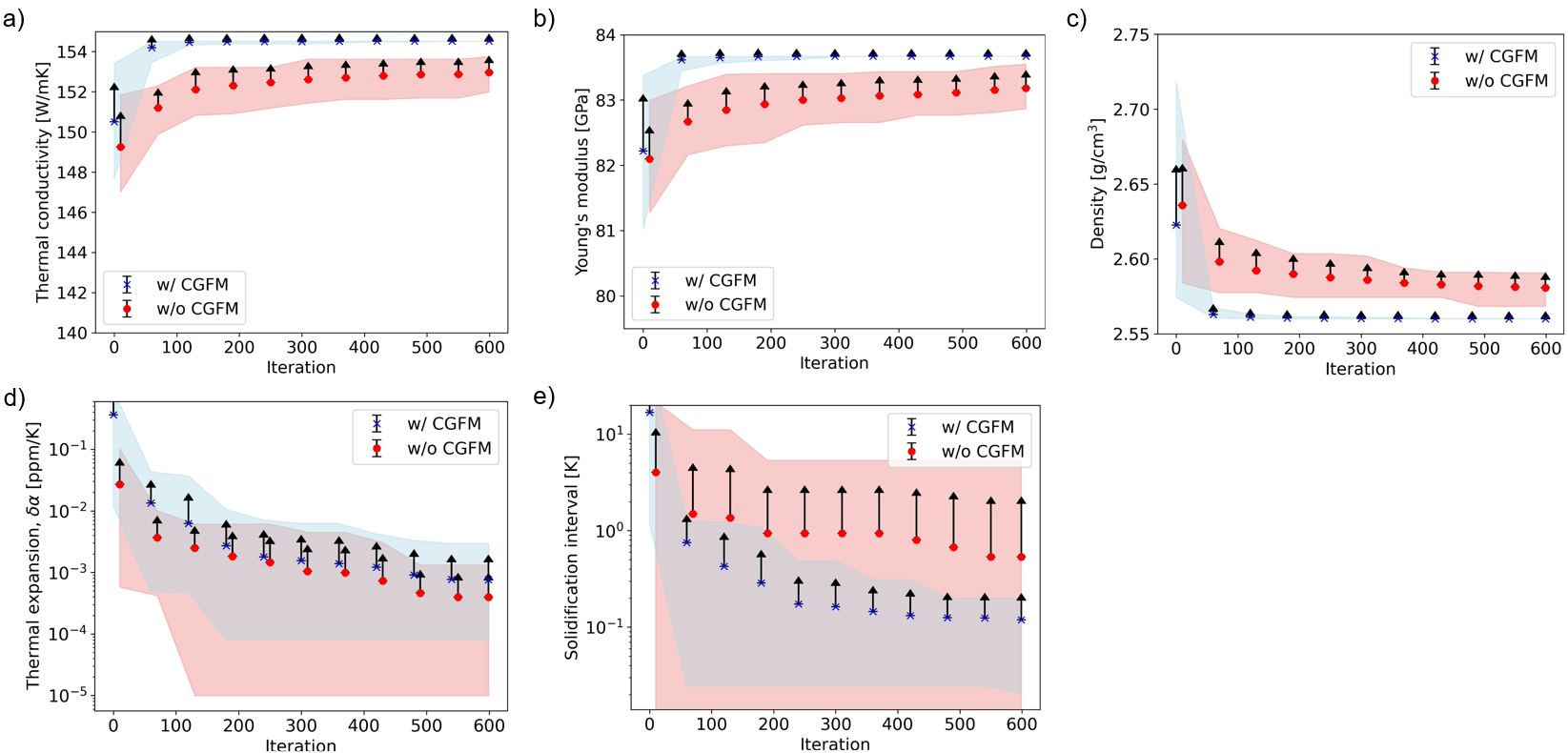}
    \caption{Performance comparison of the FM+QO procedure w/ CGFM (blue crosses) and w/o CGFM (red dots) over 600 active learning iterations implemented for single-objective optimization. a) maximization of thermal conductivity; b) maximization of Young's modulus, c) minimization of density; d) minimization of linear coefficient of thermal expansion  (absolute difference to a defined target value), $|\alpha-\alpha_0|$ with $\alpha_0=20$ ppm/K); e) minimization of solidification interval. Please note the log-scale in d and e.  
    In all plots the central markers give the mean values taken from 20 statistical replicates each started from randomized initial data. Arrows and shadings indicate the corresponding standard deviations and observed min-to-max ranges, respectively.
    }
    \label{fig_results_so_1}
\end{figure}

We investigate the single-objective variant of the FM+QO active learning approach and search for multi-phase Al-alloys with
\begin{enumerate}
    \item maximum thermal conductivity, $\kappa$,
    \item maximum Young's modulus, $E$,
    \item minimal density, $\rho$,
    \item matching thermal expansion coefficient with a target value, $\delta\alpha$, and 
    \item minimal solidification interval, $\Delta T$.
\end{enumerate}
Regarding $\delta \alpha$ we defined the target value $\alpha_0 = 20$ ppm/K, which lies rather central in the range of possible values (see details in Sec.~\ref{sec_CD_system_setup}), and to remind, the normalized fractions, $\bar f_k \in [0,1]$ with $k=1,...,N_\text{ph}$, of the $N_\text{ph}=4$ secondary phases Si, Mg$_2$Si, Al$_3$Ni and Al$_2$Cu are our model system variables.

The impact of our ansatz, referred to as \textit{w/ CGFM}, is studied by comparing its overall optimization progress against the standard approach, referred to as \textit{w/o CGFM}, which uses constraint terms as in Eq.~\ref{eq_constraint}. 
The analytical single-objective optimal solutions of the property models (see Sec.~\ref{sec_problem_setting_Prediction_models}) can be found in the Supplementary Information \ref{sec_app_best_designs}.

The five single-objective optimization cases are set up similarly. In the approach w/o CGFM ansatz, each of the four $\bar f_k$ is one-hot encoded using 50 binary variables leading to 200 binaries in total. Ultimately, this results in a 0.4 vol.\% resolution of the alloy composition. Similarly, the three constraint-guided feature variables, $\omega_k$, are one-hot encoded with each 50 binary variables following the approach w/ CGFM, which in sum consumes only 150 binary variables. 
However, the ultimate resolution of the alloy compositions can be very different. In fact, the mapping to transformed variables, $\omega_k$, allows for diverse discretizations. Its analytical expression and reverse form can be read in the Supplementary Information \ref{sec_CD_SO}, Eq.~\ref{eq_mapping_ratio_to_polar}.
Crucially, this is always in accordance with the constraint, Eq.~\ref{eq_constraint}. Here, we conveniently stay with equal-sized increments of $\omega_k$ and observe that the straight back mapping to $\bar f_k$ introduces possible alloy fractions much smaller than 0.4 vol.\%. 
The Supplementary Information \ref{sec_app_2} elaborates on this intrinsic effect.
Given the large bit numbers, we substitute a classical simulated annealing (SA) algorithm for the optimizer part in the FM+QO, i.e., here QO=SA. 
All initial data sets consist of 100 randomly selected Al-alloy designs. The algorithms w/ and w/o CGFM are started from the same initial data sets. Full details on data generation and algorithm settings can be found in the Supplementary Information \ref{sec_CD_random_des}.

The comparison of the optimization processes is shown in  Fig.~\ref{fig_results_so_1}, where the accumulated best objective values, including all previous iterations, are pointed out. The statistics results from starting the FM+QO procedure w/ and w/o the CGFM ansatz from 20 different initial data sets. The latter are the same for Fig.~\ref{fig_results_so_1}a-e. Concerning $\kappa$ (Fig.~\ref{fig_results_so_1}a, to be maximized), $E$ (Fig.~\ref{fig_results_so_1}b, to be maximized) and $\rho$ (Fig.~\ref{fig_results_so_1}c, to be minimized), the optimization clearly performs better  w/ CGFM. In particular, in both scenarios the runs w/o CGFM seem to get stuck and converge too early. Nevertheless, since even after 600 iterations the different statistical runs show rather wide spreading, we assume not a single local trap can be responsible for that. Contrarily, the different run repetitions w/ CGFM show common convergence towards the known global optima ($\kappa_{opt}=154.5$ W/mK; $E_{opt}=83.7$ GPa; $\rho_{opt}=2.56$ g/cm$^3$, see also Supplementary Information \ref{sec_app_best_designs}).

Concerning the optimization of $\delta\alpha$ and $\Delta T$, no pronounced performance differences are observed (Figs.~\ref{fig_results_so_1}d and e, respectively, on log-scale, both to be minimized). Instead, the optimizations w/ and w/o CGFM converge rather similarly.
The underlying reason lies in the specific mechanism of FM training based on the encoded constraint-guided feature variables, $\omega_k$, and in how this encoding shapes and simplifies the solution search.

To substantiate this interpretation, we first analyze the optimization scenarios and their global optima in more detail. Based on these findings, we subsequently draw conclusions regarding the strengths and limitations of the CGFM ansatz. In general, the more complex the interplay among the four phase fractions becomes, the more approximate the training of the second-order FM surrogate model tends to be. As already indicated in Sec.~\ref{sec_CGFM}, the CGFM ansatz simplifies the constrained optimization, but this simplification comes at the cost of presenting a more challenging learning task to the FM.

This trade-off can be understood by examining the feature mapping used in the CGFM formulation. A variation of an angle, $\omega_k$, necessarily induces changes in at least one of the phase fractions $\bar f_k$. Because the mapping must remain within the feasible subspace defined by $\sum_k \bar f_k = 1$, alterations of $\omega_k$ typically modify multiple fractions simultaneously. While this interdependence ensures feasibility by construction, it also amplifies the limitations of a second-order FM model. 
However, this effect is not critical when the optimum corresponds to a low-entropy solution, i.e., an alloy design dominated by a few principal constituents, especially when the variable assignments within the mapping are reshuffled during each active-learning iteration. In such cases, the benefits of unconstrained optimization can fully outweigh the approximation errors introduced by the mapping. In extreme scenarios, the CGFM ansatz may even be advantageous for the learning process itself.

This is exactly demonstrated by the $\kappa$, $E$ and $\rho$ optimizations in Figs.~\ref{fig_results_so_1}a, b and c, respectively. The optimal alloy designs are determined as Al+Al$_2$Cu, Al+Si, and Al+Mg$_2$Si, respectively, where we refer again to the Supplementary Information \ref{sec_app_best_designs} for details. These designs are extreme examples of minimum disorder, because in all three cases only a single secondary phase contributes with $\bar f_k = 1$ (i.e., $f_k = 20\%$ plus 80\% fixed Al matrix).
But because alloy designs with $\bar f_k>0.6$ are highly unlikely in our initial data sets, large individual phase fractions are unknown for the FM learning model and therefore it struggles to discover better $\kappa$, $E$ or $\rho$ designs right from the beginning of the active learning.
Instead, the optimization progress can only successively move to such designs by means of  iterative data set refinement. Thereby, we observe much weaker performance of the approach w/o CGFM ansatz. We identified two reasons for this behaviour. First, apparently the involved constraint, Eq.~\ref{eq_constraint}, complicates the search for feasible low entropy designs. 
We demonstrate this by repeating the optimization runs w/o CGFM while spending substantially more search effort, i.e., by increasing both the number of SA runs and sweeps (see Fig.~\ref{fig_results_so_1_appendix_1} of the Supplementary Information \ref{sec_optimizer_power_appendix}). We find the expected improvement of the optimization performance, but the gain saturates and cannot be scaled higher by applying more searching power to meet the performance of the approach w/ CGFM ansatz.
The second reason for the extraordinary performance of the CGFM ansatz is its distinct training data representation. Namely, w/o CGFM ansatz all $N_{ph}$ variables are relevant for learning any alloy design, even with only a single secondary phase. In contrast, for the same learning task, in the CGFM ansatz only a single variable might be relevant. Consequently, also the training in general can improve from the intrinsic constraint treatment of the CGFM ansatz, providing better exploration capabilities within less active learning iterations. A comprehensive analysis of the observed performance gains for increasing system sizes is the subject of ongoing work.

\begin{table}
\centering

\begin{tabularx}{1.0\textwidth}{ |p{1.8cm}||X|X|X|X|p{2.6cm}|  }
 \hline
 \multicolumn{6}{|c|}{\textbf{Discovered Al-alloy designs with maximal $\kappa$}} \\
 \hline
 Ansatz &  Si & Mg$_2$Si & Al$_3$Ni & Al$_2$Cu &  \textbf{$\kappa$ [W/mK]}\\
 \hline
 w/ CGFM &  0.0 (0.0\%)&  0.0 (0.0\%)& 0.0 (0.0\%) & 1.0 (20.0\%)&\textbf{154.52*} \\
 w/o CGFM &  0.0 (0.0\%) & 0.0 (0.0\%) & 0.2 (4.0\%) & 0.8 (16.0\%) &\textbf{153.75} (154.52) \\ 
 \hline
 \end{tabularx}
 \vspace{0.1cm}

\begin{tabularx}{1.0\textwidth}{ |p{1.8cm}||X|X|X|X|p{2.8cm}|  }
 \hline
 \multicolumn{6}{|c|}{\textbf{Discovered Al-alloy designs with maximal $E$}} \\
 \hline
 Ansatz &  Si & Mg$_2$Si & Al$_3$Ni & Al$_2$Cu &  \textbf{$E$ [GPa]}\\
 \hline
 w/ CGFM &  1.0 (20.0\%)&  0.0 (0.0\%)& 0.0 (0.0\%) & 0.0 (0.0\%)&\textbf{83.671*} \\
 w/o CGFM &  0.98 (19.6\%) & 0.0 (0.0\%) & 0.0 (0.0\%) & 0.02 (0.4\%) &\textbf{83.517} (83.671) \\ 
 \hline
 \end{tabularx}
 \vspace{0.1cm}

\begin{tabularx}{1.0\textwidth}{ |p{1.8cm}||X|X|X|X|X|  }
 \hline
 \multicolumn{6}{|c|}{\textbf{Discovered Al-alloy designs with minimum $\rho$}} \\
 \hline
 Ansatz &  Si & Mg$_2$Si & Al$_3$Ni & Al$_2$Cu &  \textbf{$\rho$ [g/cm$^3$]}\\
 \hline
 w/ CGFM &  0.0 (0.0\%)& 1.0 (20.0\%)& 0.0 (0.0\%)& 0.0 (0.0\%) & \textbf{2.56*} \\
 w/o CGFM &  0.08 (1.6\%) & 0.9 (18.0\%) & 0.0 (0.0\%) & 0.02 (0.4\%) &\textbf{2.574} (2.56) \\ 
 \hline
 \end{tabularx}
 \vspace{0.1cm}

\begin{tabularx}{1.0\textwidth}{ |p{1.8cm}||X|X|X|X|X|  }
\hline
 \multicolumn{6}{|c|}{\textbf{Discovered Al-alloy designs with $\delta\alpha$ closest to target}} \\
 \hline
 Ansatz &  Si & Mg$_2$Si & Al$_3$Ni & Al$_2$Cu &  \textbf{$\delta\alpha$ [ppm/K]}\\
 \hline
 w/ CGFM &  0.342 (6.84\%)& 0.018 (0.35\%)& 0.483 (9.65\%)& 0.158 (3.15\%) & \textbf{8e-5} (3e-6) \\
 w/o CGFM &  0.32 (6.4\%) & 0.24 (4.8\%) & 0.34 (6.8\%) & 0.1 (2.0\%) &\textbf{1e-5*} \\ 
 \hline
 \end{tabularx}
 \vspace{0.1cm}

\begin{tabularx}{1.0\textwidth}{ |p{1.8cm}||p{2.6cm}|X|X|X|X|  }
 \hline
 \multicolumn{6}{|c|}{ \textbf{Discovered Al-alloy designs with minimum $\Delta T$}} \\
 \hline
 Ansatz &  Si & Mg$_2$Si & Al$_3$Ni & Al$_2$Cu &  \textbf{$\Delta T$ [K]}\\
 \hline
 w/ CGFM & 0.6398 (12.797\%) & 0.224 (4.48\%) & 0.135 (2.71\%) & 0.001 (0.02\%) & \textbf{0.02083*}  \\
 w/o CGFM &  0.64 (12.8\%) & 0.2 (4.0\%)& 0.14 (2.8\%)& 0.02 (0.4\%) &\textbf{0.0*} \\ 
 \hline

\end{tabularx}
\caption{Optimal single-objective Al-alloy designs discovered from the 20 statistical runs of Fig.~\ref{fig_results_so_1} comparing FM+QO w/o and w/ CGFM ansatz. Top to bottom: thermal conductivity, Young's modulus, density, linear coefficient of thermal expansion (i.e., $\delta \alpha = | \alpha - \alpha_0|$), and solidification interval.
The alloy designs are described in terms of the phase fractions of Si, Mg2Si, Al3Ni and Al2Cu, $\bar f_k\in[0,1]$, where $\bar f_k=1$ is here equivalent to the maximum phase contribution of 20 \% (reminding fixed 80 \% Al matrix). For interpretation, translated values to vol.[\%] are given in brackets.
Following the CGFM ansatz, the constraint-guided feature variables are encoded based on equally-spaced discretization, which results in non-equally spaced increments when translated back to the $\bar f_k$. In the approach w/o CGFM ansatz the $\bar f_k$ are directly encoded giving equally spaced increments. The corresponding objective values are each given in the last column. The true optimal values, considering the discrete search space, are given in brackets or indicated by stars.
}
\label{table_so_opt_res}
\end{table}

For $\delta\alpha$ and $\Delta T$, the best alloy designs found by FM+QO render much more diverse results.
Tab.~\ref{table_so_opt_res} shows the best designs of the 20 runs obtained w/ and w/o CGFM (taken from Fig.~\ref{fig_results_so_1}), respectively.
We further analysed the solution density around the optimal alloy design by brute force search directly based on our property models (with details given in the Supplementary Information \ref{sec_brute_force_appendix}).
In particular, solution density close to the ground state in cases of $\delta\alpha$ and $\Delta T$ is much higher compared to $\kappa$, $E$ or $\rho$. Furthermore, close to the optimal designs regarding $\delta\alpha$ and $\Delta T$ we find rather homogeneous compositions of all four phases. In particular, this results from the chosen target value $\alpha_0=20$ ppm/K, which can be approached by many different alloy designs. Moreover, in the simplified model of $\Delta T$ only the fraction of eutectic Si is relevant leading to many possible degenerate ground states.
In such scenarios like $\delta\alpha$ and $\Delta T$, the FM training is harder when encoding the feature variables of the CGFM ansatz, and its weakness against the direct encoding of the fraction variables (w/o CGFM) hides its strengths.

The bottom line is that FM+QO w/ CGFM exhibits superior performance in identifying alloy designs characterized by only a few dominant secondary phases. Compared to FM+QO w/o CGFM, it explores regions of the search space that are not represented in the initial data set more efficiently. However, when targeting optimal designs with more evenly distributed component ratios, the benefits of the CGFM ansatz have to compensate for the more complicated FM training.
Finally, it should be noted, that across all scenarios our CGFM ansatz consistently accelerates the active learning process when measured in real time, owing to QUBO model simplification, as demonstrated in detail in the Supplementary Sec.~\ref{sec_CGFM_speedup}.

\subsubsection{Multi-objective optimization with DDTS ansatz}
\label{sec_mo_opt}

We now employ the multi-objective FM+QO active learning approach to systematically evaluate the performance of our DDTS ansatz. Therefore we maintain the Al-alloy model system already studied above having a fixed 80 vol.[\%] Al-matrix and described in terms of normalized fraction variables, $\bar f_k\in[0,1]$ with $k=1,...N_\text{ph}$, of the $N_\text{ph}=4$ phases: eutectic Si, Mg$_2$Si, Al$_3$Ni and Al$_2$Cu. 
Further, we keep the standard FM+QO without CGFM and apply one-hot encoding to translate each of the discretized $\bar f_k$  to a number of  $N_\text{bits} = 25$ binary variables. 
Supplementary Information \ref{sec_CD_MO} holds further computational details.

Given the inherent conflict of optimizing $\kappa$, $E$ and $\rho$ simultaneously, these material properties are well-suited to demonstrate multi-objective Pareto optimization algorithms. Although we focus the following discussion on this scenario, other combinations of the five objectives show similar behaviour and all results can be found in Supplementary Information \ref{sec_app_1}.

\begin{figure}
    \centering
    \includegraphics[width=\textwidth]{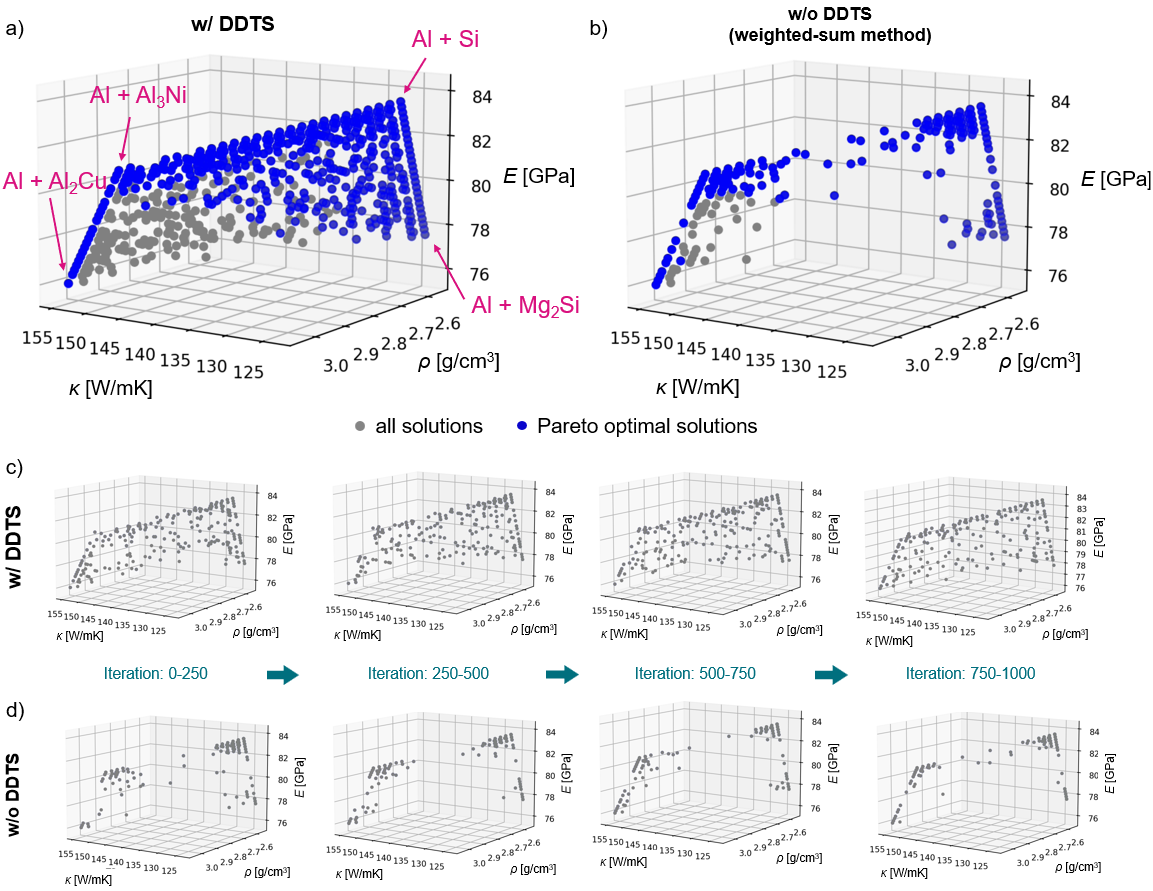}
    \caption{Multi-objective optimization results with the FM+QO method based on either the DDTS ansatz (w/ DDTS) or the standard weighted-sum scalarization (w/o DDTS). Thermal conductivity ($\kappa$), Young's modulus ($E$) and mass density ($\rho$) span the objective space. a and b) Solutions found within 1000 iterations w/ and w/o DDTS, respectively. Gray points are all QUBO solutions (best alloy designs) found. (These can appear multiple times, the substituted random designs are not illustrated.) The blue points illustrate the 3D Pareto front (non-dominating solutions). DDTS ansatz delivers a homogeneous distribution of Pareto optimal alloy designs across the entire front, which is in stark contrast to the ansatz w/o the DDTS. According to our property models, the alloy designs in the Pareto "corners" are given by the four possible binary Al alloys : Al+Si, Al+Mg$_2$Si, Al+Al$_3$Ni and Al+Al$_2$Cu.
    c) Temporal progress of solution search corresponding panel a. Left to right: solutions sampled within iterations 0-250, 250-500, 500-750 and 750-1000, respectively. d) Similar plots corresponding to panel b.}
    \label{fig_results_mo_1}
\end{figure}

The DDTS workflow is depicted in the upper part of Fig.~\ref{fig_DDTS}.
Now let us compare the generated Pareto fronts based on the DDTS to the results obtained from the FM+QO approach with standard \textit{weighted-sum scalarization} (shown in the lower part of Fig.~\ref{fig_DDTS}).
For convenience, we refer to both ansatzes as FM+QO \textit{w/ DDTS} and \textit{w/o DDTS} in the following.
Pareto results from the multi-objective FM+QO algorithm are presented in the 3D objective space spanned by $\kappa$, $E$ and $\rho$ in Fig.~\ref{fig_results_mo_1}. Fig.~\ref{fig_results_mo_1}a and b refer to optimization runs w/ and w/o DDTS, respectively. The active learning process is stopped after completing 1000 iterations. The preference weights, $w_i$, for setting the different scalarizations in both ansatzes are randomly sampled for each active learning iteration in order to generate the entire Pareto front. 
The comparison of the Pareto fronts in Figs.~\ref{fig_results_mo_1}a and b demonstrates the advantage of our DDTS ansatz against the standard weighted-sum scalarization.
The blue dots give each the obtained set of Pareto optimal solutions called \textit{Pareto front}. Also the remaining gray dots are alloy designs proposed by the QO, however, those are not in the final Pareto front.
The DDTS ansatz delivers a homogeneous distribution of Pareto optimal alloy designs across the entire Pareto front. To prove this, the Supplementary Information \ref{sec_pareto_brute_force} provides the exact Pareto front based on brute force search. 
Apparently, the FM+QO w/ DDTS is capable to generate Pareto optimal designs with equal probability. In this regard, the ansatz w/o DDTS performs poorly. In Fig.~\ref{fig_results_mo_1}b, we observe the found Pareto solutions concentrate on the peripheral regions.
The time-resolved generation of QUBO solutions, i.e., new optimal design candidates, is visualized in Fig.~\ref{fig_results_mo_1}c and d, respectively.
The temporal evolution demonstrates the ability of the ansatz w/ DDTS to homogeneously improve the front over time. Moreover, it shows that throughout the entire active learning process the ansatz w/o DDTS samples only solutions from peripheral regions. 
In fact, the planar parts of the Pareto front are inaccessible when solution search is based on the weighted-sum scalarization. 
Only solutions around the corresponding single-objective optima (extreme preference weights): Al+Al$_2$Cu, Al+Si and Al+Mg$_2$Si can be returned, as well as in the forth "corner" of the front (around Al+Al$_3$Ni). Noteworthy,  the latter binary composition is not a single-objective optimum in our model but it is responsible for a local convexity which can be straightly accessed by the convex approach.

We conclude this section by noting that the FM+QO with DDTS ansatz performs exceptionally well for data-driven multi-objective optimization problems. It clearly outperforms standard weighted-sum scalarization and enables the treatment of inherently non-convex QUBO problem settings. In contrast to the weighted-sum method, our DDTS ansatz enables the identification of Pareto-optimal Al-alloy designs even in non-convex regions of the Pareto front, ranging here from binary to quaternary compositions.

Future work will involve systematic studies comparing FM+QO w/DDTS to established non-convex multi-objective optimizers, such as evolutionary algorithms or particle swarm optimization.

\subsection{Implications}

\label{sec_conclusion}

We see significant potential in QUBO-based quantum optimization for real-world applications. However, significant research challenges remain -- both in advancing quantum hardware and in developing algorithmic approaches capable of achieving quantum advantage. 

Contributing to the development of tailored algorithms, we present two straightforward and easily implementable methods customized to data-driven, real-valued black-box optimization problems.

\textbf{Constraint-guided feature mapping (CGFM):} This approach enables the
reformulation of QUBO problems with equality constraints by transforming the problem into specially oriented coordinates. This coordinate transformation reduces the solution space to the lower-dimensional subspace of feasible solutions, effectively eliminating equality constraints. As a result, the QUBO formulation becomes more compact, parameterization is simplified, and subsequent optimization, whether performed via quantum annealing (QA), quantum approximate optimization algorithms (QAOA), or classical heuristics, is significantly accelerated.
    
\textbf{Data-driven Tchebycheff scalarization (DDTS):} This pre-processing method supports efficient QUBO-based multi-objective optimization. Inspired by the Tchebycheff scalarization method, which is well-established for non-convex, multi-objective problems outside the QUBO domain, we adapt the technique to the data-driven QUBO setting. Our formulation guarantees the generation of optimal solutions across the entire Pareto front including non-convex regions. This is a notable improvement over current scalarization strategies in QUBO, which often fail to explore large regions of the front.

We apply both methods to a complex, multi-dimensional, multi-objective optimization task from materials research, namely the design of a multi-phase alloy considering a hypothetical microstructure at µm-scale. More realistic microstructures considering actual descriptors for engineering materials will be considered in the next developments.
Looking ahead, we plan to explore the CGFM approach in conjunction with quantum hardware, particularly quantum annealers. These platforms may still offer sufficient qubit capacity to model complex composite ratios effectively. Additionally, a systematic evaluation of CGFM under different QUBO encoding schemes remains an open question. For the latter issue we will expand our studies by investigating the alternative options of regular binary and domain-wall encoding. We consider such extensions as promising avenues for future research.

\section{Materials and Methods}
\label{sec_methods}

\subsection{Multi-phase alloy model}
\label{sec_problem_setting}

Our optimization problem describes a hypothetical Al-alloy with four components (phases), which can be found in the microstructure of many cast Al-alloys at µm-level. Besides the Al-matrix these are:  eutectic Si, Mg$_2$Si, Al$_3$Ni and Al$_2$Cu. The task is to vary the fractions [\%] of the $N_{\text{ph}}=4$  phases, $f_k$ with $k\in N_{\text{ph}}$, in order to maximize $\kappa$ and $E$ as well as to minimize $\rho$, $ \delta \alpha$ and $\Delta T$. Note further, that with 
\begin{align}
\delta \alpha = | \alpha - \alpha_0 |
\label{eq_dalpha}
\end{align}
we minimize the absolute value difference of the thermal expansion, $\alpha$, with respect to a given target value $\alpha_0$. The content of the aluminum matrix is fixed to 80 \% vol.~fraction, which a realistic value for near eutectic Al-Si alloy.

\subsubsection{Thermal and mechanical objectives}

\label{sec_problem_setting_Prediction_models}

We are aware of the generally complex 3D-architecture that  phases can form in cast Al-Si alloys \cite{Asghar2009,Asghar2010,Tolnai2013}, however, we keep a very simplified description of the alloy, as shown in Fig.~\ref{fig:discretization}a (left). Thereby, we assume that all four phases can be considered as spherical isotropic particles which are homogeneously distributed in the aluminum matrix and do not interact with each other. 

Following these assumptions, we employ the Mori-Tanaka model \cite{Benveniste1986} to predict the thermal conductivity of the alloy according to
\begin{align}
\kappa = \kappa_\text{Al} \frac{1 + 2 Q}{1-Q}
\label{eq_k}
\end{align}
with 
\begin{align}
Q = \sum_{k=1}^{N_{\text{ph}}}   \frac{\kappa_k - \kappa_\text{Al}}{2 \kappa_\text{Al} + \kappa_k } f_k.
\end{align}
Conveniently, the $k$-th phase fraction is introduced in relative units here, i.e. $f_k \in [0,0.2]$, which we use hereafter.
The $\kappa_k$ and $\kappa_\text{Al}$ indicate the thermal conductivity of phase $k$ and the Al-matrix (see parameterization in Sec.~\ref{sec_CD_system_setup}).

The Young’s modulus is estimated in an iterative manner by the Hill model \cite{Bunge2000}. Hill's model suites well to our morphology assumptions and the resulting $E$ values lie within the upper and lower boundaries given by Voigt  \cite{Voigt1910} and Reuss \cite{Reuss1929} for fibre-like and plate-like morphology, respectively. The Hill model is initialized by determining
\begin{equation}
\begin{aligned}
E_\text{Hill} &= \frac{1}{2} \big(E_\text{Reuss} + E_\text{Voigt} \big) \\
S_\text{Hill} &= \frac{1}{2} \big(E_\text{Reuss}^{-1} + E_\text{Voigt}^{-1} \big)
\end{aligned}
\label{Hill_init}
\end{equation}
with the Young's moduli, $E_\text{Reuss}$ and  $E_\text{Voigt}$, according to the Reuss and Voigt models,
\begin{equation}
\begin{aligned}
E_\text{Reuss} &= \Big[  \frac{f_\text{Al}}{E_\text{Al}} + \sum_{k=1}^{N_{\text{ph}}} \frac{f_k}{E_k}  \Big]^{-1} \\
E_\text{Voigt} &= f_\text{Al} E_\text{Al} + \sum_{k=1}^{N_{\text{ph}}} f_k E_k .
\end{aligned}
\label{E_r_v}
\end{equation}
The $E_\text{Al}$ and $E_k$ in Eqs.~\ref{E_r_v} indicate the Young's modulus of the matrix and the $k$-th phases, respectively (see parameterization in Sec.~\ref{sec_CD_system_setup}). Subsequently, the initial values, Eq.~\ref{Hill_init}, are utilized to solve the equations
\begin{equation}
\begin{aligned}
E_\text{Hill,'} &= \frac{1}{2} \big(E_\text{Hill} + S_\text{Hill}^{-1} \big) \\
 S_\text{Hill,'} &= \frac{1}{2} \big(E_\text{Hill}^{-1} + S_\text{Hill} \big) 
\end{aligned}
\label{eq_E}
\end{equation}
self-consistently until convergence is found in terms of $E_\text{Hill} =  S_\text{Hill}^{-1} $. 

The mass density of the alloy, $\rho$, is the weighted arithmetic mean considering the density of the aluminum matrix, $ \rho_\text{Al}$, and the other phases, $\rho_k$,
\begin{align}
\rho = f_\text{Al} \rho_\text{Al} + \sum_{k=1}^{N_{\text{ph}}} f_k \rho_k.
\label{eq_rho}
\end{align}
In particular, aerospace and automotive industries are interested in lightweight Al-alloys to reduce fuel consumption. 

The coefficient of thermal expansion indicates the volumetric change of a material as a function of temperature. In particular, the linear coefficient of thermal expansion, $\alpha$, is defined by the ratio of the relative change in length and the applied temperature difference. Here, we utilize the simple and frequently used Turner model \cite{Turner1946}, which follows as
\begin{align}
\alpha = \frac{f_\text{Al} \alpha_\text{Al} K_\text{Al} + \sum_{k=1}^{N_{\text{ph}}} f_k \alpha_k K_k }{f_\text{Al} K_\text{Al} + \sum_{k=1}^{N_{\text{ph}}} f_k K_k  }
\label{eq_alpha}
\end{align}
and is based on the bulk modulus of the aluminum matrix, $K_\text{Al}$, and the $k$-th phase, $K_k$  (see parameterization in Sec.~\ref{sec_CD_system_setup}).
Eq.~\ref{eq_alpha} assumes isostrain conditions, i.e., it is suitable for predicting $\alpha$ in alloys with continuous reinforcements, which, though beyond the assumptions introduced above, describes our hypothetical microstructure reliably.
Instead of optimizing $\alpha$ directly, we intend to minimize  the absolute difference, $\delta\alpha$ (Eg.~\ref{eq_dalpha}) to a predefined target value. Controlling the thermal expansion of Al alloys can be important in multi-component systems that operate at different temperatures.

Finally, we introduce the solidification interval, $\Delta T$, that gives an indication of the castability of an alloy. As a rule of thumb, it can be considered that the castability decreases with increasing solidification intervals. For a very first estimate, we simplify our picture of the multi-phase alloy and derive $\Delta T$ from the Al-Si binary phase diagram, in particular, the temperature difference between $L$-phase and $\alpha+$Si-phase. To do that analytically, we apply a linear fitting to the phase boundaries in the range of 0-20 \% Si. The resulting linear expression approximately models $\Delta T$ only depending on the silicon fraction, $f_\text{Si}$, 
\begin{equation}
\Delta T= 
    \begin{cases}
        -M_1 f_\text{Si} + N_1 & \text{if } f_\text{Si} < 0.128 \\
        0 & \text{if } f_\text{Si} = 0.128 \\
        M_2 f_\text{Si} - N_2  & \text{if } f_\text{Si} > 0.128
    \end{cases}
\label{eq_deltaT}
\end{equation}
with slopes $M_1$ and $M_2$ and shifts $N_1$ and $N_2$ (see Sec.~\ref{sec_CD_system_setup} for details). Silicon content of 12.8 \% approximately marks the eutectic point, i.e. $\Delta T=0~K$, in the Al-Si binary alloy.

\subsubsection{Material parameters}
\label{sec_CD_system_setup}

The Eqs.~\ref{eq_dalpha}-\ref{eq_deltaT} are utilized to determine values for our five target properties: thermal conductivity ($\kappa$), Young's modulus ($E$), density ($\rho$), linear coefficient of thermal expansion ($\alpha$) and the solidification interval ($\Delta T$) within our Al-alloy design model (see Sec.~\ref{sec_problem_setting}). 
For the phases Si, Mg$_2$Si, Al$_3$Ni and Al$_2$Cu, we utilize the materials parameters given in Tab.~\ref{table_param}. Note that we directly maximize $\kappa$ (see Eq.~\ref{eq_k}) and $E$ (see Eq.~\ref{eq_E}) as well as directly minimize $\rho$ (see Eq.~\ref{eq_rho}) and $\Delta T$ (see Eq.~\ref{eq_deltaT}). The thermal expansion coefficient, however, is optimized to match best a defined target, i.e., our third optimization goal regarding $\alpha$  is to minimize the absolute difference $|\alpha-\alpha_0|$ with $\alpha$ determined by Eq.~\ref{eq_alpha} and the target value defined as $\alpha_0 = 20$ ppm/K. $\alpha_0$ is chosen without particular purpose but to be in the mid of range of possible $\alpha$ values resulting from the variation of the four phase fractions. This range stretches from $\alpha=18.2$ ppm/K for (80 \% Al + 20 \% Si) to $\alpha=21.9$ ppm/K for (80 \% Al + 20 \% Al$_2$Cu).

\begin{table}
\centering
\begin{tabular}{ |c||c|c|c|c|c|  }
 \hline
 \multicolumn{6}{|c|}{Precipitate phase types and material properties} \\
 \hline
 Phase & $\kappa$ [W/m K] & $E$ [GPa] & $\rho$ [g/cm$^3$] & $\alpha$ [ppm/K] & $K$ [GPa] \\
 \hline
 Al (matrix) & 164$^\text{a}$ & 70 \cite{Barnoush2012} & 2.7 & 23.6 & 70 \cite{Mo2023} \\
eutectic Si & 15 \cite{Zhang2023} & 163$^\text{b}$ & 2.3 & 2.6 \cite{Watanabe2004,Middelmann2015} & 97$^\text{b}$\\
 Mg$_2$Si & 11 \cite{Martin1972,Tani2008} & 116 \cite{Lan2019,Tani2008} & 2.0 & 7.5 \cite{Lan2019} & 59 \cite{Yu2010} \\
 Al$_3$Ni &104 \cite{Mo2023} &140 \cite{Fukui1994,Hertzberg1965, Xiao2025} & 4.0& 13.0 \cite{Mo2023}& 110 \cite{Mo2023,Xiao2025} \\
Al$_2$Cu &120 \cite{Zhang2016,Wei2016} & 105 \cite{Wang2021,Tian2017} &4.4 & 17.0 \cite{Chen2010,Lokker2000} & 97 \cite{Jin2024} \\
 \hline
\end{tabular}
\caption{Materials properties of the five phases considered in the Al alloy, as introduced in Sec.~\ref{sec_problem_setting} ($\kappa$: thermal conductivity, $E$: Young's modulus, $\rho$: density, $\alpha$: linear coefficient of thermal expansion and $K$: bulk modulus).
The parameters are required for computing the optimization objectives of the Al-alloy along the prediction models described in  Eqs.~\ref{eq_dalpha}-\ref{eq_deltaT}. References from literature related to the parameters are stated. \\ 
$^\text{a}$ Please note we assume thermal conductivity of the aluminum matrix, $\kappa_\text{Al}=164$ W/mK. The value lies below the value of 237 W/mK of pure aluminum due to inherent minor impurities. Density and elastic modulus are less effected by those. \\
$^\text{b}$ The bulk modulus of silicon, $K_\text{Si}$, is approximately determined based on $K_\text{Si}=E_\text{Si}/(3-6v_\text{Si})$ for isotropic materials with the Young's modulus, $E_\text{Si}=163$ GPa, and the Poisson ratio, $v_\text{Si} = 0.22$, of silicon. The value $E_\text{Si}$ is chosen as a reasonable estimate for eutectic silicon.}
\label{table_param}
\end{table}

\begin{figure}
    \centering
    \includegraphics[width=0.6\textwidth]{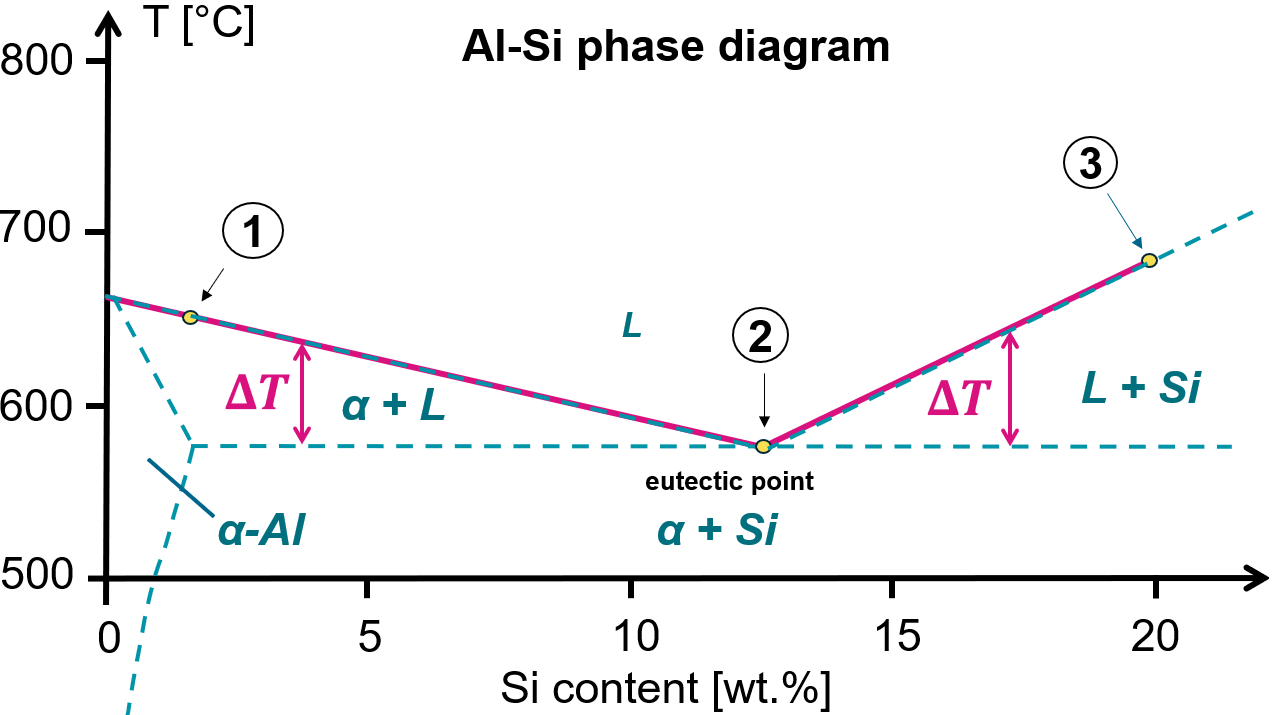}
    \caption{Phase diagram of binary Al-Si alloy. The solidification interval, $\Delta T$, describes the temperature difference between the liquid and solid phase ($L$ and $\alpha + \text{Si}$ phase, respectively) and depends on Si content. Here, the range from 0-20 \% Si is of interest. According to our assumption, we require the indicated 3 points for linearly fitting $\Delta T$. 1: (1.6, $T_S$+75), 2: (12.8, $T_S$), 3: (20, $T_S$+100) with the solidification temperature $T_S=577$.}
    \label{fig_deltaT_fitting}
\end{figure}

Furthermore, we implement a simple approximation based on the binary Al-Si system to determine the solidification interval, $\Delta T$. The analytic expression in Eq.~\ref{eq_deltaT} describes a linear fitting model of the temperature difference between the boundaries of the liquid and solid phase. For this purpose, we approximate the multi-phase alloy with the solidification interval of the Al-Si binary composition. This assumption crucially allows for straightly determining the fitting parameters $M_1$, $M_2$, $N_1$ and $N_2$ in Eq.~\ref{eq_deltaT} based on the phase boundaries in the Al-Si phase diagram. A schematic representation of the phase diagram and the fitted boundaries are given in Fig.~\ref{fig_deltaT_fitting}. We are aware of the limitation of our simplification but it helps us to demonstrate the methodology used for the optimization of a multi-component system adding a further variable, namely the solidification interval. A more realistic approach to actually discover a new alloy composition could, e.g., consider determining the solidification interval using the CALPHAD method or by experimental studies of the real alloy compositions. 
The three  points indicated in the phase diagram are used to fit
$\Delta T [K]= - M_1 f_\text{Si} + N_1 $ with $M_1=669.6$ and $N_1=85.7$ for $f_\text{Si} \in [0,0.128]$, as well as $\Delta T [K]=  M_2 f_\text{Si} - N_2 $ with $M_2=1388.9$ and $N_2=177.8$ for $f_\text{Si} \in [0.128,0.2]$ (see the magenta lines in Fig.~\ref{fig_deltaT_fitting}). For simplicity, we  exclude the $\alpha$-Al phase from our consideration (in the range $f_\text{Si} \in [0,0.016]$). Point (2) in Fig.~\ref{fig_deltaT_fitting} is the eutectic point of $\Delta T =0$ K. The $\Delta T$ is to be minimized in our optimization scenarios, therefore we expect FM+QO to return ideal alloys with $f_\text{Si} = 0.128$ (i.e., $\bar f_\text{Si} = 0.64$), as far as the introduced search space discretization allows.

\subsection{Model encoding}
\label{sec:encoding}

\subsubsection{Discretization of phase fractions}

The microstructure of our hypothetical Al-alloy is illustrated in Fig.~\ref{fig:discretization}a. The design is defined by the fractions of the phases, $f_k$, with $k=1,..., N_\text{ph}$, which serve our model as variables in the range [0, 0.2], reminding the fixed Al matrix content of 0.8. For the FM+QO method, this continuous variable space must be translated into binary variable space. In the most straightforward manner, an equally-spaced discretization of the [0, 0.2] range can be introduced for this purpose \cite{Arai2023, Yarkoni2022}. The mesh is applied to all variables and determines the resolution of the design model. For instance, using a discretization with 9 mesh points would allow describing fractions in multiples of 0.02 (i.e., 2 \%). The discretized variable axes are finally converted into $N_\text{bits}$ binary variables, $x_i$, with $i=1,..., N_\text{bits}$ (see Sec.~\ref{sec_enc_details}). Though more types of 0-1 encoding are well known from literature, we focus on one-hot encoding types \cite{Yarkoni2022, Dominguez2023}.  The total number of required variables is determined by multiplying the number of bits, $N_\text{bits}$, with the number of phases, $N_\text{ph}$.

\subsubsection{Encoding of phase fractions}
\label{sec_enc_details}

To utilize FM+QO the discretized variable ranges of the phase fractions must be further encoded in binary variables. In preparation, we first conveniently introduce \textit{normalized} phase fractions, $\bar f_k \in [0,1]$, according to
\begin{equation} 
   f_k = L + (U-L) \bar f_k
   \label{eq_relative_fractions}
\end{equation}
with $L$ and $U$ indicating the lower and upper boundaries of the $f_k$ variable range, respectively. Please note, in our model we identify $L=0$ and $U=0.2$.
Given a certain number of bits, $N_{\rm bits}$, available for the encoding of a single normalized variable, $\bar f_k$, we write
\begin{equation} \label{encoding}
   \bar f_k = \sum_{i=1}^{N_{\rm bits}} \alpha_i x_i^{(k)}
\end{equation}
Here, $x_i^{(k)} \in \{0,1\}$ is the $i$-th binary variable (or bit) used to encode the fraction $\bar f_k$. The factors $\alpha_i$ depend on the used type of encoding. In general,  $\alpha_i$ could be $k$-dependent as well, however, here we process each $\bar f_k$ similarly.

For the one-hot encoding, we have
\begin{align}
 \alpha_i = i \cdot \Delta,
\end{align}
with $i=1,..., N_\text{bits}$, where $\Delta = 1 / N_\text{bits}$ is the increment size resulting from the discretization of normalized fractions, $\bar f_k$. Note that to obtain the absolute increment size we would simply have to multiply this value by $U-L$ (i.e., here with a factor of 0.2). The name one-hot encoding refers to the fact that, for this type of encoding, at most one bit can be 1 while the rest has to be 0.
In contrast to the classical one-hot encoding, we encode $\bar f_k=0$ by the zero bit-string $x_i = 0$ for all $i$ with the advantage of saving one bit. In this way, intuitively, each bit corresponds to a specific contribution to $\bar f_k$. If bit $i$ is set to 1, a part of $i/\Delta$ contributes to the total sum in Eq.~\ref{encoding}. Consequently, the set of representable values spans uniformly the interval $[0,1]$.

\subsection{FM+QO active learning}
\label{sec_fmqo_al}

In general, the FM+QO methodology works through an iterative process of step-wise improvements repeating \cite{Kitai2020}:
\begin{enumerate}
    \item data-based FM modeling (FM),
    \item QUBO optimization (QO), 
    \item enlargement of the data set with the validated solutions
\end{enumerate}
until some form of convergence is reached (see Fig. 2). According to the state-of-the-art \textit{weigted-sum scalarization}, this straightly generalizes to multi-objective optimization tasks with $m$ objectives as follows:
Let the data set consist of vectors, $\vec {x}$, describing the system in terms of binary variables, $x_i \in \small\{ 0,1\small\}$, and the corresponding $m$ optimization target properties, $y^{(1)}, ...y^{(m)}$, i.e., the $d$-th data point follows  as $(\vec{x}_d,y^{(1)}_d, ...y^{(m)}_d)$. 
Then, one FM is trained independently for each of the $m$ objectives. Details on the FM modeling analytics can be found in the Supplementary Information \ref{sec_app_3}.

\begin{figure}
    \centering
    \includegraphics[width=\textwidth]{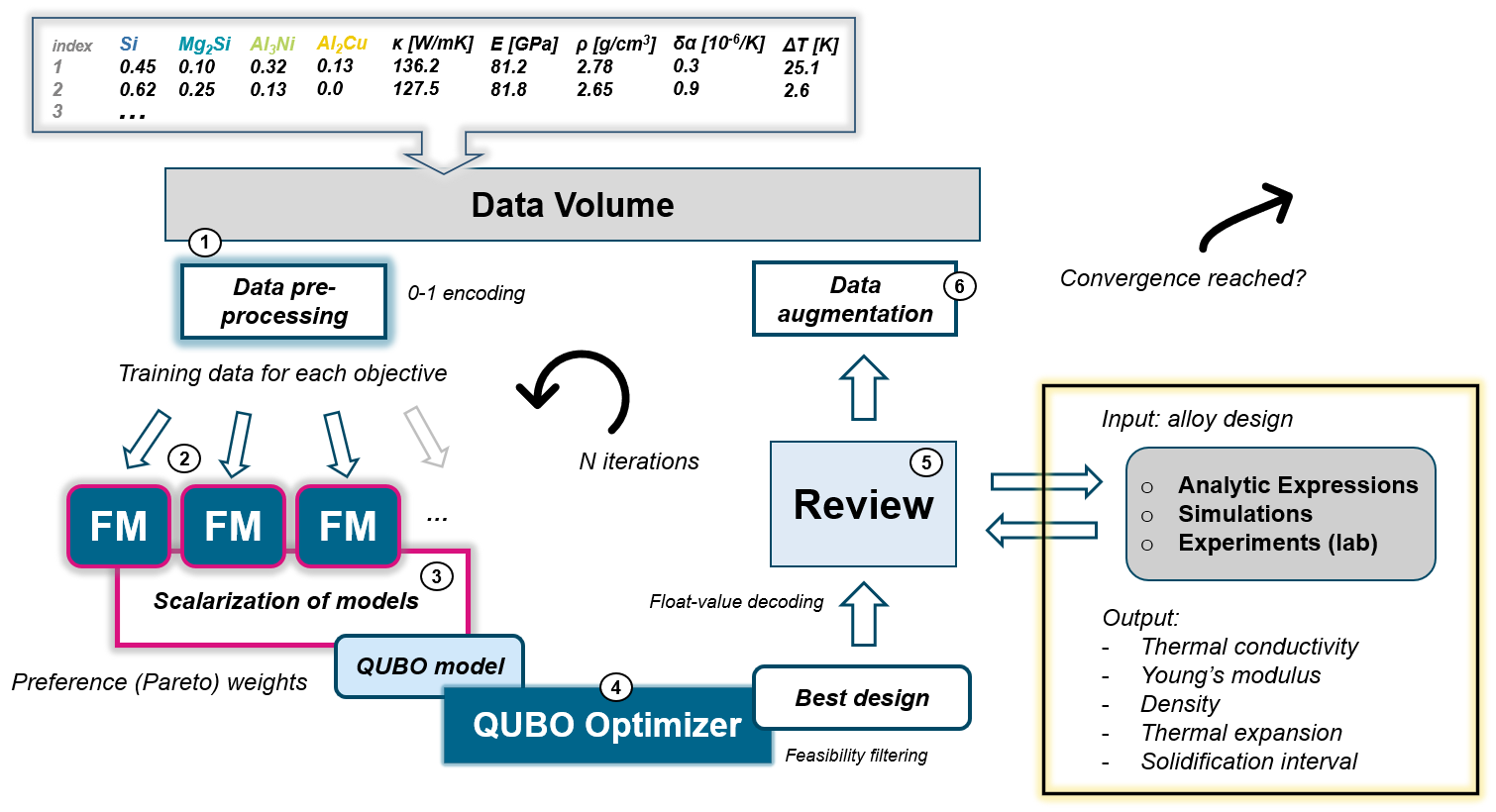}
    \caption{The multi-objective implementation of the active learning FM+QO method based on recent state-of-the-art QUBO scalarization, here applied for optimizing the microstructure of a hypothetical Al-alloy in terms of normalized phase fractions, $\bar f_k \in [0,1]$  with $k=1,...,N_\text{ph}$(see Eq.~\ref{eq_relative_fractions}). Each iteration starts from encoding and pre-processing the existing data set (1). A factorization machine (FM) is trained for regression of each of the given thermal and mechanical objectives (2). The FMs are scalarized to merge the surrogate models into a single one (3). The resulting QUBO model is solved for searching a new optimal  design (4). Finally, the objective's values of the found design are reviewed by means of a validation source, e.g., analytic expressions, simulations or experiments (5). Here, we make use of Eqs.~\ref{eq_dalpha}-\ref{eq_deltaT}. This gives a new data point which is added to the data set for improving the modeling in the next iteration (6). Note that the FM+QO pipeline based on our new DDTS ansatz differs in steps (1) - (3), see Fig.~\ref{fig_DDTS}.}
    \label{fig:fmqo_scheme}
\end{figure}

The resulting QUBO models, $f^{(p)}(\vec{x})$, with objective index $p=1,...m$, read
\begin{align}
f^{(p)}(\vec{x}) = \vec{x}^T \hat{Q}^{(p)} \vec{x} ,
\label{eq_QUBO_single_obj}
\end{align}
based on the learned matrices $\hat{Q}^{(p)}$, the multi-objective model finally follows from merging all $f^{(p)}(\vec{x})$ into a single QUBO,
\begin{align}
g(\vec{x}) = \sum_{p=1}^{m} w_{p} \vec{x}^T \hat{Q}^{(p)} \vec{x}  +\lambda C(\vec{x}).
\label{eq_QUBO}
\end{align}
In principle, choosing specific weights, $w_p$, where usually $\sum_p w_p = 1$, controls the merger preferences regarding the objectives. Finally, system required constraints $C(\vec{x})$ must be added to the QUBO with penalty strength factor $\lambda$, to support feasible solutions. Please note that here we conveniently suppressed constraints which additionally arise for specific encodings like the one-hot type and instead refer the interested reader to Eq.~\ref{eq_constr_one_hot} in Sec.~\ref{sec_CD_SO}.
After solving $g(\vec{x})$ with QO, the learned model is validated and improved. Practically this is realised by determining the "true" objective values for the QO solution and by adding this new, validated data point to the data set. This review process might be facilitated by simulations, experiments, or available analytic formulae. By refining the data set in this manner, the FM model improves iteratively.

In our specific use case, each initial data point represents a random alloy design. Its search space vector, $\vec x_d$, describes a concatenation of $N_\text{ph}$ vectors, $\vec{x}^{(k)} = (x_1^{(k)},...,x_{N_\text{bits}}^{(k)})$, with $x_i^{(k)} \in \small\{ 0,1\small\}$ and $k=1,...N_\text{ph}$, obtained by one-hot encoding the normalized phase fraction variables, $\bar f_k$ (see Sec.~\ref{sec:encoding}). Moreover, each data point holds the thermal and mechanical property values, $y_d^{(p)}$ with $p\in \small\{\kappa,E,\rho,\delta\alpha,\Delta T\small\}$, introduced in Sec.~\ref{sec_problem_setting} and determined by querying the prediction models in Eqs.~\ref{eq_dalpha}-\ref{eq_deltaT}. Importantly, these expressions are also our validation sources to review the predicted properties of the optimal Al-alloy design in each iteration (step 5 in Fig.~\ref{fig:fmqo_scheme}).

\section{Acknowledgements}
We acknowledge the financial support of the DLR Quantum Computing Initiative through the project QuantiCoM (\url{https://qci.dlr.de/quanticom/}).

\section{Competing interests}
T.~P., D.~M., D.~B. and E.~B. are named inventors on the pending patent application DE102025118818.1, and all authors are named inventors in the pending patent application DE102025147173.8, both filed by the German Aerospace Center (DLR), related to the results described in this work.

\newpage

\bibliographystyle{naturemag}
\bibliography{main}

\end{document}


\maketitle

\keywords{QUBO-based quantum optimization, multi-objective optimization, optimization under constraints, Tchebycheff features, factorization machine}

\listofabbreviation{CGFM: Constraint-guided feature mapping, DDTS: Data-driven Tchebycheff scalarization, QUBO: Quadratic unconstraint binary optimization, FM: Factorization machine, QO: QUBO optimization, QA: Quantum annealing, QAOA: Quantum approximate optimization algorithm, SA: Simulated annealing, $\kappa$: Thermal conductivity, $E$: Young's modulus, $\rho$: density, $\alpha$: Linear coefficient of thermal expansion, $\Delta T$: Solidification interval}

\tableofcontents

\section{Computational details}

\subsection{Factorization machine regression model} 
\label{sec_app_3}

Our QUBO optimization approaches are aided by the machine learning method called factorization machine (FM) \cite{Rendle2010}. The method provides a model $f^{\text{FM}}(\vec{x})$ which directly works on binary variables, $x_i$, and it attributes interactions among the $x_i$ up to second order,
\begin{align}
f^{\text{FM}}(\vec{x}) = w_0 + \sum_{i=1}^{N} w_i x_i + \sum_{i<j}^{N} w_{ij} x_i x_j,
\label{eq_FM}
\end{align}
with the constant bias, $w_0$, the weight of each variable, $w_i$, and the weights of pairwise interaction among the variables, $w_{ij}$.
Essentially, the $N^2$ parameters $w_{ij}$ are not directly trained but determined from a factorization ansatz according to
\begin{align}
w_{ij} = \langle  \mathbf{v}_i, \mathbf{v}_j \rangle.
\end{align}
By training instead the vectors $\mathbf{v}_i\in \mathbb{R}^{k}$ allows for more compact modeling based on only $N\cdot k$ parameters. The $\mathbf{v}_i$ can be interpreted as embeddings of the $x_i$ in a latent space, where the factorization rank, $k$, defines the dimensionality of the latent space. Then it is, the closer $x_i$ and $x_j$ are embedded in latent space, the stronger their interaction is weighted in the trained model. Concerning the $\mathbf{v}_i$ as the $i$-th column of a matrix $V\in \mathbb{R}^{k\times N}$, it follows $w_{ij} = (V^T V)_{ij}$, which is similar to a matrix factorization ansatz and inspired the name of the method.
Finally, we drop the overall bias, $w_0$, and rewrite Eq.~\ref{eq_FM} to matrix product form
\begin{align}
f^{\text{FM}}(\vec{x}) = \vec{x}^T \hat{Q} \vec{x},
\end{align}
where $\hat Q$ denotes the QUBO matrix of the FM model (see also Eq.~\ref{eq_QUBO_single_obj} in the main article), with diagonal elements, $Q_{ii} = w_i$, and off-diagonal elements, $Q_{ij} = w_{ij}$, with $i<j$.

\subsection{Single-objective optimization with CGFM ansatz}
\label{sec_CD_SO}

For properly comparing the FM+QO progress w/ and w/o CGFM, the optimization runs are provided with the same initial data and, as far as possible, run under the same parameterization of the FM regression method and SA optimizer.
Here we give further information on the computational implementation.

The randomized initial data set generation follows the procedure declared in Sec.~\ref{sec_CD_random_des}.  Each active learning iteration starts with pre-processing the corresponding data set in order to improve the FM model training. Concerning the target properties we apply Z-score scaling for better performance. The four phase fraction values, $\bar f_k \in [0,1]$, are one-hot encoded, where our implementation also involves the binary string of only zeros encoding $\bar f_k=0.0$. Hence, the penalty term to be added in order to penalize all solutions in the search space which do not comply to the encoding reads,
\begin{align}
C^\text{oh}(\vec{x}) = \sum_{k=1}^{N_\text{ph}}  C^\text{oh}_k(\vec{x}).
\label{eq_constr_one_hot}
\end{align}
It decomposes into $N_\text{ph}$ equivalent terms corresponding to each a different phase fraction variable
\begin{align}
C^\text{oh}_k(\vec{x}) = \big( \sum_{i=1}^{N_\text{bits}} x_i^{(k)} \big)  \big( \sum_{i=1}^{N_\text{bits}} x_i^{(k)} - 1 \big),
\end{align}
where it is $C^\text{oh}_k(\vec{x}) \ge 0$. The $x_i^{(k)}$ denote the $i$-th binary variable (i.e., one of a number of $N_\text{bits}$ bits) introduced to encode $\bar f_k$. The expression minimizes to $C^\text{oh}_k(\vec{x}) = 0$ for $\sum_i^{N_\text{bits}} x_i^{(k)} \in \small[0,1\small]$, i.e., for solutions which comply with our encoding. 
When making use of the CGFM ansatz, we similarly one-hot encode the $N_\text{ph}-1$ "synthetic" contraint-guided coordinates, $\omega_k$.

The intention of the CGFM ansatz is to eliminate the constraints arising due to model system specifics (see Sec.~\ref{sec_CGFM} of the main article). Here, we have to enforce feasible alloy designs expressed by a constraint condition of type
\begin{align}
 \sum_{k=1}^{N_\text{ph}} \bar f_k  = 1.
\label{eq_constraint_SI}
\end{align}
The latter can be intrinsically obeyed when implementing the CGFM mapping $f(\bar f_1,...,\bar f_{N_\text{phm}}) = \omega_1, ..., \omega_{N_\text{ph}-1}$, i.e., $f: \mathcal{R}^{N_\text{ph}} \rightarrow \mathcal{R}^{N_\text{ph}-1}$, according to
\begin{equation}
\label{eq_mapping_ratio_to_polar}
\begin{aligned}
\omega_1 &= \text{arctan2}\Big( \big(\sum_{k=2}^{N_\text{ph}} \bar f_k \big)^{1/2}, \bar f_1^{1/2} \Big) \\
\omega_2 &= \text{arctan2}\Big( \big(\sum_{k=3}^{N_\text{ph}} \bar f_k \big)^{1/2}, \bar f_2^{1/2} \Big) \\
&... \\
\omega_{N_\text{ph}-1} &= \text{arctan2}\Big( \bar f_{N_\text{ph}}^{1/2} , \bar f_{N_\text{ph}-1}^{1/2} \Big). 
\end{aligned}
\end{equation}
As can be proved straightly, the given coordinate transformation intrinsically fulfills the condition in Eq.~\ref{eq_constraint_SI}.
The reverse mapping $f: \mathcal{R}^{N_\text{ph}-1}  \rightarrow \mathcal{R}^{N_\text{ph}}$ reads 
\begin{equation}
\label{eq_mapping_polar_to_ratio}
\begin{aligned}
\bar f_1 &= C  \big( \cos(\omega_1) \big)^2 \\
\bar f_2 &= C  \big( \sin(\omega_1) \cos(\omega_2) \big)^2\\
\bar f_3 &= C  \big( \sin(\omega_1) \sin(\omega_2) \cos(\omega_3) \big)^2\\
&... \\
\bar f_{N_\text{ph}} &= C \big( \sin(\omega_1) \cdots \sin(\omega_{N_\text{ph}-1}) \cos(\omega_{N_\text{ph}-1}) \big)^2,
\end{aligned}
\end{equation}
where the radial variable is here fixed to the parameter $C$. Its value is directly determined to $C=1.0$ according to Eq.~\ref{eq_constraint_SI}. Please note further that we shuffle the variable assignment within both mappings for each active leaning iteration to improve the FM model training.

Then for reliable training of the FM the total data set is split into sub sets utilized for training, validation and testing purpose. Thereby, we utilized the open python package fastFM \cite{Bayer2016} and the alternating least square algorithm.
For some training instances, especially in the early active learning iterations based on a small data set, we experienced strong dependence of the training performance on the FM hyperparameters. Therefore, we optimize the FM training as a simple black-box with the goal to increase the testing score. The best parameter set of the standard deviation (used to initialize the model) and L2 penalty weights (for linear and quadratic coefficients) is searched within a Optuna \cite{Optuna2019} study of 20 repetitions. The FM factorization rank is fixed to 6, and the each training runs for the maximum number of 2000 epochs.

The system and encoding related constraints are not part of the training, thus before the QUBO model, $f(\vec{x})=\vec{x}^T \hat{Q}^{(p)} \vec{x}$, obtained by the FM is solved, it must be attributed with all needed penalty terms. Without the CGFM ansatz the condition in Eq.~\ref{eq_constraint_SI} must be translated to the constraint term 
\begin{align}
C(\vec{x}) = \Big( \sum_{k=1}^{N_\text{ph}} \sum_{i=1}^{N_\text{bits}} \alpha_i x_i^{(k)} -  1 \Big)^2,
\label{eq_constraint_encoded}
\end{align}
with encoding dependent bit weights  $\alpha_i$, and added to the QUBO together with the above one-hot constraint, Eq.~\ref{eq_constr_one_hot},
\begin{align}
f'(\vec{x})=\vec{x}^T \hat{Q}^{(p)} \vec{x} + \lambda C(\vec{x}) + \gamma C^\text{oh}(\vec{x}).
\label{eq_full_QUBO_onehot}
\end{align}
This step also involves the task to find their proper scaling factors $\lambda$ and $\gamma$, respectively. Both must be balanced against each other and against the $\hat{Q}^{(p)}$ term. After normalizing all three terms in Eq.~\ref{eq_full_QUBO_onehot}, we found good performance with $\lambda=650.0$ and $\gamma=1.0$. Please note that normalization means scaling in order to achieve that the largest absolute value of each term gives 1.0.
Contrarily, when implementing the CGFM ansatz instead, only the one-hot encoding constraint $C^\text{oh}(\vec{x})$ is relevant, which generally makes finding good parametrization easier as only a single constraint term must be balanced against $\hat{Q}^{(p)}$.

The QUBO model is then solved by means of SA, where we implement the algorithm of the D-Wave Ocean software suite \cite{dwave_ocean_docs}. The number of SA runs and sweeps per optimization is fixed at 1000 and 3000, respectively (analogously for the approach w/ and w/o CGFM ansatz). The annealing schedule is automatically adjusted to the QUBO, where we employ the Ocean internal  \textit{\_default\_ising\_beta\_range()} sampler routine, which is based on the maximum and minimal effective bias per Ising spin variable, in order to assure rapid equilibration at the beginning (hot temperature) and a small rate of excitations at the end (low temperature).
Next, the \textit{true} material properties of the single best found candidate alloy design are validated in a review process based on the analytical predictor models described in Eqs.~\ref{eq_dalpha}-\ref{eq_deltaT} in the main article. In the last step of each active learning iteration the found design with its true properties is added to the data set.
This refinement of the data set requires repeating the pre-processing for the next active learning iteration.

If the QUBO optimization delivers a solution which already have been found previously, the corresponding known alloy design is replaced by a new feasible one determined randomly (see Sec.~\ref{sec_CD_random_des}).

\subsection{Multi-objective optimization with DDTS ansatz}
\label{sec_CD_MO}

In the following, computational details on our multi-objective FM+QO approach implementations  w/ and w/o DDTS ansatz  are given.

The initial data set of 500 designs is generated following the second procedure described in Sec.~\ref{sec_CD_random_des}. The first step of each active learning iteration for the FM+QO w/ and w/o DDTS is the pre-processing of the data set, where the same techniques are applied as stated in Sec.~\ref{sec_CD_SO}. Following the DDTS ansatz, next, the 3d objective space vectors, $(y_d^{(\kappa)}, y_d^{(E)},y_d^{(\rho)})$, of each data point are transformed to a single objective value, $\hat{y}_d$, according to Eq.~\ref{eq_DDTS_2} in the main article. The corresponding utopian points $u_{\kappa}$, $u_{E}$ and $u_{\rho}$ can be based on available system knowledge. However, in a more generic manner and assuming the case of no former knowledge, we empirically take  $u_{\kappa} = 1.1 \max \small\{ y_d^{(\kappa)} \small\}$ and  $u_{E} = 1.1 \max \small\{ y_d^{(E)} \small\}$ as well as $u_{\rho}= 1.1 \min \small\{ y_d^{(\rho)} \small\}$, respectively, which places the utopia always somewhat ahead of the best known objective values (reminding the applied Z-score scaling). Further, the weight vector, $\vec{w}$, is randomly chosen. Following the DDTS ansatz, next, the FM model is trained on the transformed data set, $(\vec{x}_d, \hat{y}_d)$. The automated hyperparameter search described in Sec.~\ref{sec_CD_SO} is used. Finally, the obtained regression QUBO model is given to the QO optimizer unit for searching the best alloy design. Concerning the FM+QO w/o DDTS instead three FM models have to be trained for individual prediction of $\kappa$, $E$, and $\rho$, where the fixed parameters are chosen identical to the approach w/ DDTS and the automatically adjusted parameters are optimized under the same conditions. Then, the three learned QUBOs are finally merged making use of the weighted sum method and the scalarized QUBO is fed into the optimization.
To close the iteration, only the single best solution from the optimization is validated in our review process (in accordance with Sec.~\ref{sec_CD_SO}).

\subsection{Random design generation}
\label{sec_CD_random_des}

In order to initialize the FM+QO method, we set up random data sets of Al-alloy designs. The Al-matrix content of our alloys is fixed to 80 \%, and the remaining 20 \% is variable in terms of the  $N_\text{ph}=4$ secondary phase contributions. For convenient modeling we work with normalized phase fractions $\bar f_k \in [0,1]$ (see Eq.~\ref{eq_relative_fractions} in the main article Sec.~\ref{sec:encoding}). A single data point in the set comprises all phase fractions, $\bar f_k$ with $i\in N_\text{ph}$, and the target properties to be optimized. In particular, the data set consists only of feasible materials, i.e., each data point always obeys $\sum_{k=1}^{N_\text{ph}} \bar f_k =  1$.

\begin{figure}
    \centering
    \includegraphics[width=0.5\textwidth]{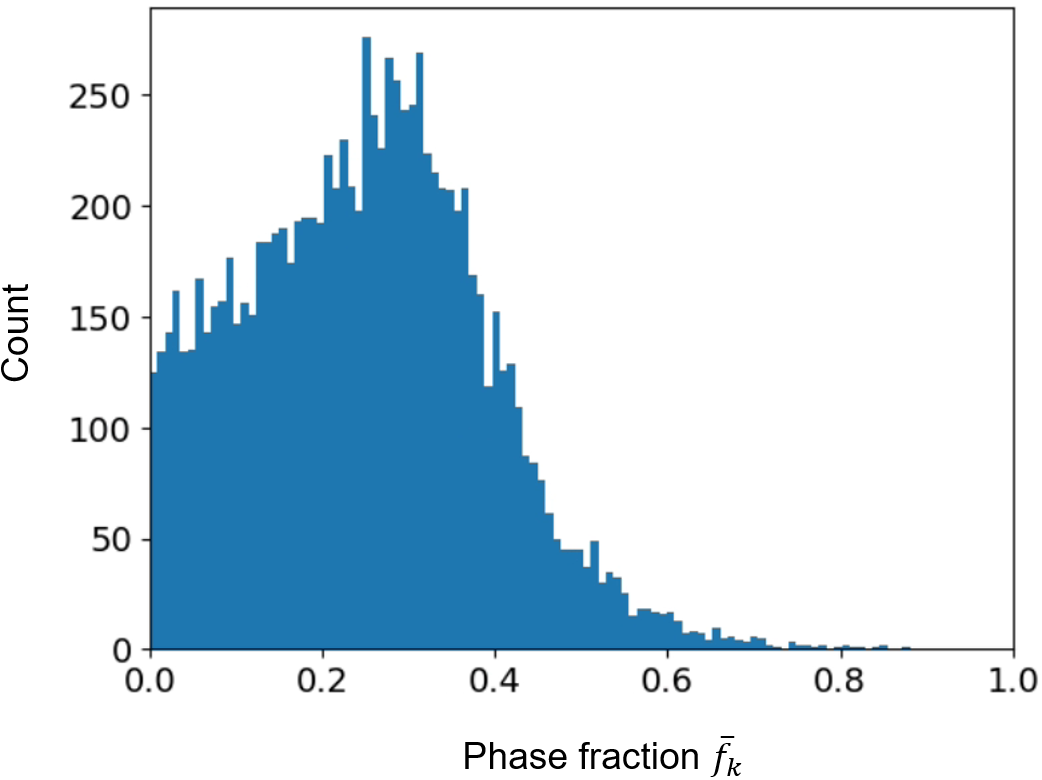}
    \caption{Distribution of normalized fraction values, $\bar f_k \in [0,1]$, from sampling equally distributed random alloy designs with $N_\text{ph}=4$ secondary phases (excluding the Al-matrix) as implemented for the single-objective FM+QO studies. The random sampling accounts for the systematic constraint, $\sum_{k=1}^{N_\text{ph}} \bar f_k = 1.0$. The histogram shows that phase fractions of $\bar f_k > 0.6$ are highly unlikely sampled. }
    \label{fig_ratio_dist}
\end{figure}

For the single-objective FM+QO studies in Sec.~\ref{sec_CGFM} of the main article, we generate random alloy designs by drawing first $N_\text{ph}-1$ random numbers, $a_l \sim \mathcal{U}( 0, 1) $ with $l\in N_\text{ph} - 1$, which are then sorted by value and listed to $A = \text{sort}( \small\{ a_l \small\})$. Next, $N_\text{ph}$ fractions, $\tilde f_k$, are taken from $A$ according to, $\tilde f_k = A_k - A_{k-1}$ with $k\in N_\text{ph}$ and where we add $A_0 = 0$ and $A_{N_\text{ph}}= 1$. The last step generated a shuffled assignment of $\tilde f_k$ to the $N_\text{ph}$ constituents, i.e., the final random design is $\bar f_k = \tilde f_{\pi(k)}$ with $k=1,...,N_\text{ph}$.
A histogram over the individual $\bar f_k$ generated in this way is given in Fig.~\ref{fig_ratio_dist} for $N_\text{ph}=4$. According to the generator, ratios of $\bar f_k> 0.6$ are very unlikely. Please note, that for increasing $N_\text{ph}$ this effect becomes even more crucial.

For the multi-objective FM+QO studies in Sec.~\ref{sec_DDTS} of the main article, we proceeded the following. In order to properly study the optimization performance of the DDTS ansatz independent on the performance of the QUBO model regression, the FM training is supported by introducing a much larger data set and a random design generator with a bias towards larger phase fractions, $\bar f_k$. Instead of the above random alloy generator scheme, we apply now the following. To start with, a first fraction $\tilde f_1$ is drawn uniform randomly from the full interval, i.e., $\tilde f_1 \sim \mathcal{U}( 0, 1) $. The remaining values, $\tilde f_k$ with $k\in 2,..., N_\text{ph}-1$, are step-wise randomly generated from a reduced interval according to, $\tilde f_k \sim \mathcal{U}(0, 1 - \tilde f_{k-1}) $. The last fraction then remains with $\tilde f_{N_\text{ph}} = 1- \tilde f_{N_\text{ph}-1}$. Finally, the fraction values, $\tilde f_k$, are shuffled again to generate the random design, $\bar f_k = \tilde f_{\pi(k)}$ with $k=1,...,N_\text{ph}$.

\section{Side effect of discretizing the synthetic variables} 
\label{sec_app_2}

\begin{figure}
    \centering
    \includegraphics[width=0.65\textwidth]{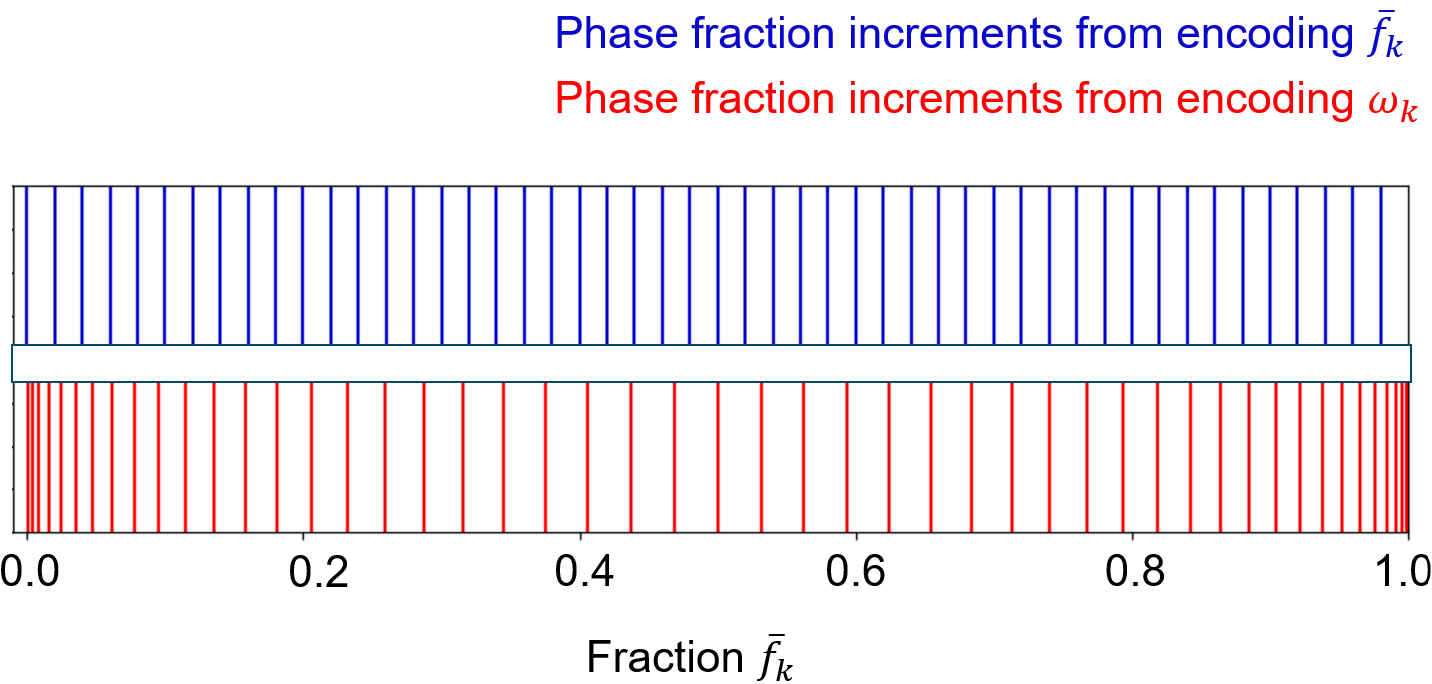}
    \caption{Illustration of different increments along the phase fractions, $\bar f_1$, depending on the utilized discretization. 50 grid points are utilized per coordinate. The blue lines in the upper part indicate 50 equal-sized increments from directly discretizing the fractions, $\bar f_k$. The red lines in the lower part illustrate the internal effect when our CGFM ansatz is applied, i.e., when discretizing the synthetic contraint-guided  variables, $\omega_k$, instead. Then, the fraction increments obtained by Eq.~\ref{eq_mapping_polar_to_ratio} are of different size and are concentrated at the interval boundaries.}
    \label{fig_polar_disc}
\end{figure}

Working in the subspace spanned by the synthetic variables, $\omega_k$, brings with an interesting side effect. In fact, according to the mapping, Eq.~\ref{eq_mapping_polar_to_ratio}, encoding equal-sized increments along $\omega_k$ finally translates into differently sized increments of the normalized phase fraction variables, $\bar f_k \in [0,1]$. This intrinsic effect is illustrated in Fig.~\ref{fig_polar_disc}. The upper blue lines indicate 50 increments of $\bar f_1$ obtained from the standard approach, where the fractions are directly encoded and the entire search space is discretized into increments of equal volume. The lower red lines plot the corresponding increments of $\bar f_1$ when applying the similar discretization to the $w_i$ variables instead. As shown, it allows for much more precisely resolving the fractions, $\bar f_k$, at its interval boundaries (i.e., in the shown example for $\bar f_1$ close to 0.0 and 1.0). Moreover, concerning the $\bar f_k$ with $k>1$ in Eq.~\ref{eq_mapping_polar_to_ratio}, we recognize dependencies on two or respectively more $\omega_k$. The non-equal grid meshing of the $\bar f_k$ dynamically adjusts in this way in order to ensure that the alloy design stays within the subspace of feasible solutions.
We interpret this as an advantage of the CGFM ansatz in scenarios where major as well as very tiny fractions of secondary phases are crucial.

\section{FM+QO: Performance depending on optimizer power}
\label{sec_optimizer_power_appendix}

\begin{figure}
    \centering
    \includegraphics[width=0.5\textwidth]{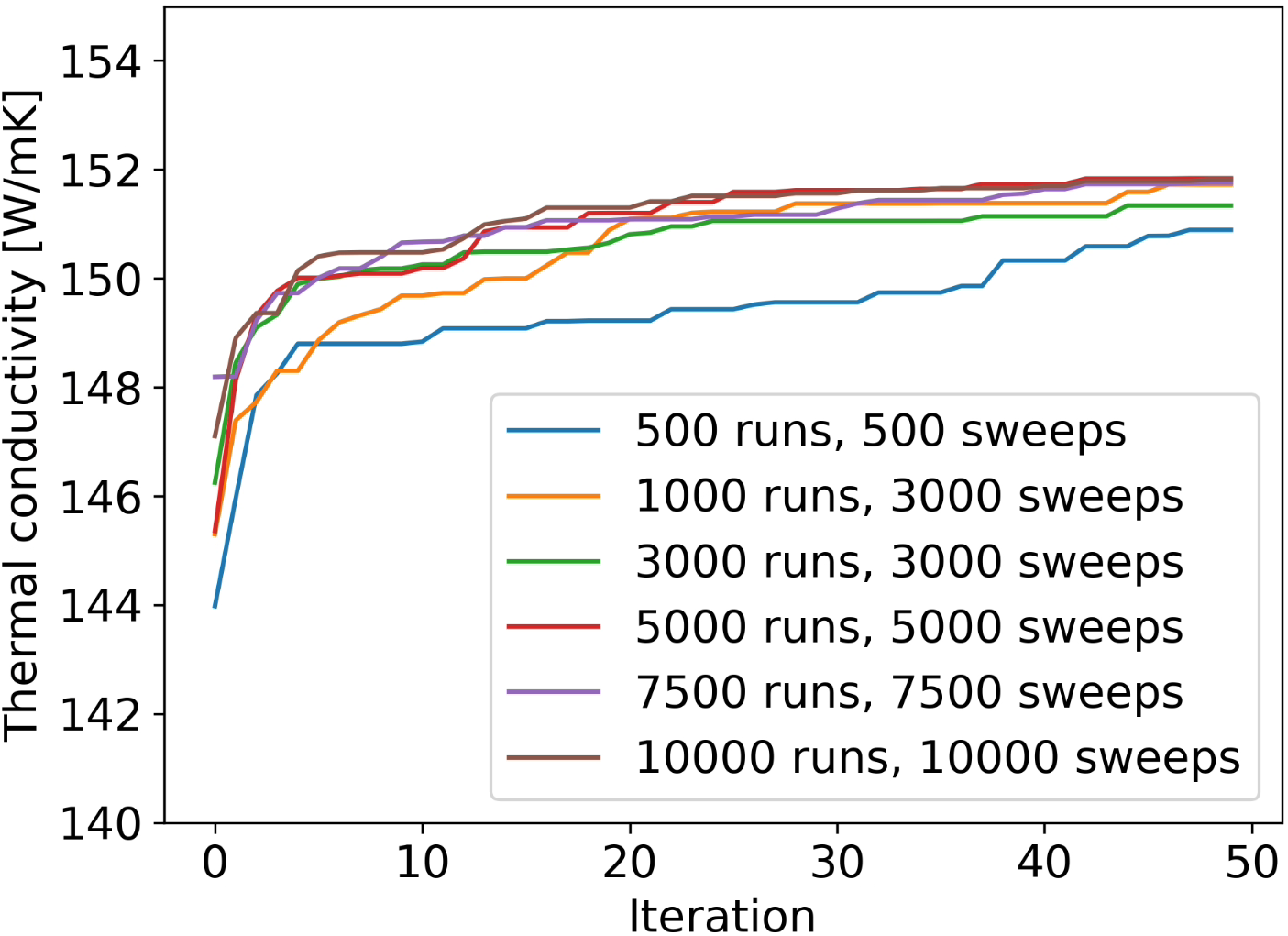}
    \caption{Dependence of the FM+QO progress on the applied QO power (see discussion in the main article Sec.~\ref{sec_so_opt}). Here QO=SA and the searching power is scaled in terms of number of simulated annealing runs (i.e., repetitions) and sweeps (i.e., variation epochs) as given in the legend. The optimization scenario is similar to the red line in Fig.~\ref{fig_results_so_1}a (maximization of $\kappa$; FM+QO w/o CGFM ansatz). Here, the initial phase of 50 FM+QO active learning iterations is recorded, where each line is the average over 20 statistical FM+QO repetitions started from different data sets.}
    \label{fig_results_so_1_appendix_1}
\end{figure}

The optimization progress of the FM+QO method depends on the power of the applied optimizer (QO). Supplementing the discussion in the main article Sec.~\ref{sec_so_opt}, Fig.~\ref{fig_results_so_1_appendix_1} demonstrates this dependence for the maximization of $\kappa$ w/o the CGFM ansatz. The initial progress during 50 active learning iterations is illustrated. The comparison documents clearly the saturation of the impact on the performance by increasing the QO computational effort.

\section{FM+QO: Real-time speed-up with the CGFM ansatz}
\label{sec_CGFM_speedup}

The optimization performance of the FM+QO method w/ CGFM ansatz is studied in the main article Sec.~\ref{sec_so_opt}. Let us here discuss the real-time acceleration achieved by the CGFM ansatz in detail. 

The run time of a single FM+QO iteration is largely determined by two factors: the duration of the training of the FM to create the regression model and the duration of the subsequent QO (here SA). The duration of the FM training depends on training algorithm, model size, the applied rank of the model and the number of training epochs required. The FM rank shall not be further considered, as it was always fixed to 6 throughout our experiments w/ and w/o CGFM.
On the one hand, the model size, $MS$, in the standard case w/o CGFM is given by $MS=N_{\rm ph} \cdot N_{\rm bits}$ with $N_{\rm ph}$ continuous fraction variables and $N_{\rm bits}$ binary variables for their encoding. On the other hand, for the model w/ CGFM ansatz, $MS=(N_{\rm ph}-d) \cdot N_{\rm bits}$, where $d$ denotes the dimension of the constraint, that is the number of variables eliminated. In our case, $d=1$. Our training is based on an alternating least square algorithm \cite{Bayer2016}, and scales linearly with $MS$. Consequently, the duration of a training epoch also scales linearly in $N_{\rm ph}$ (with a fixed number $N_{\rm bits}$), i.e., the method w/ CGFM has an advantage over using the standard encoding w/o CGFM, which scales with $N_{\rm ph}/(N_{\rm ph}-1)$.
Besides, there is obviously the influence of the number of necessary training epochs. However, that influence is extremely dependent on the respective data set or use case, and it is difficult to predict. In our experiments, the possible number of training epochs was always limited to a maximum value of 2000, but a certain deviation from the  $N_{\rm ph}/(N_{\rm ph}-1)$  scaled advantage is to be expected.

\begin{figure}
    \centering
    \includegraphics[width=\textwidth]{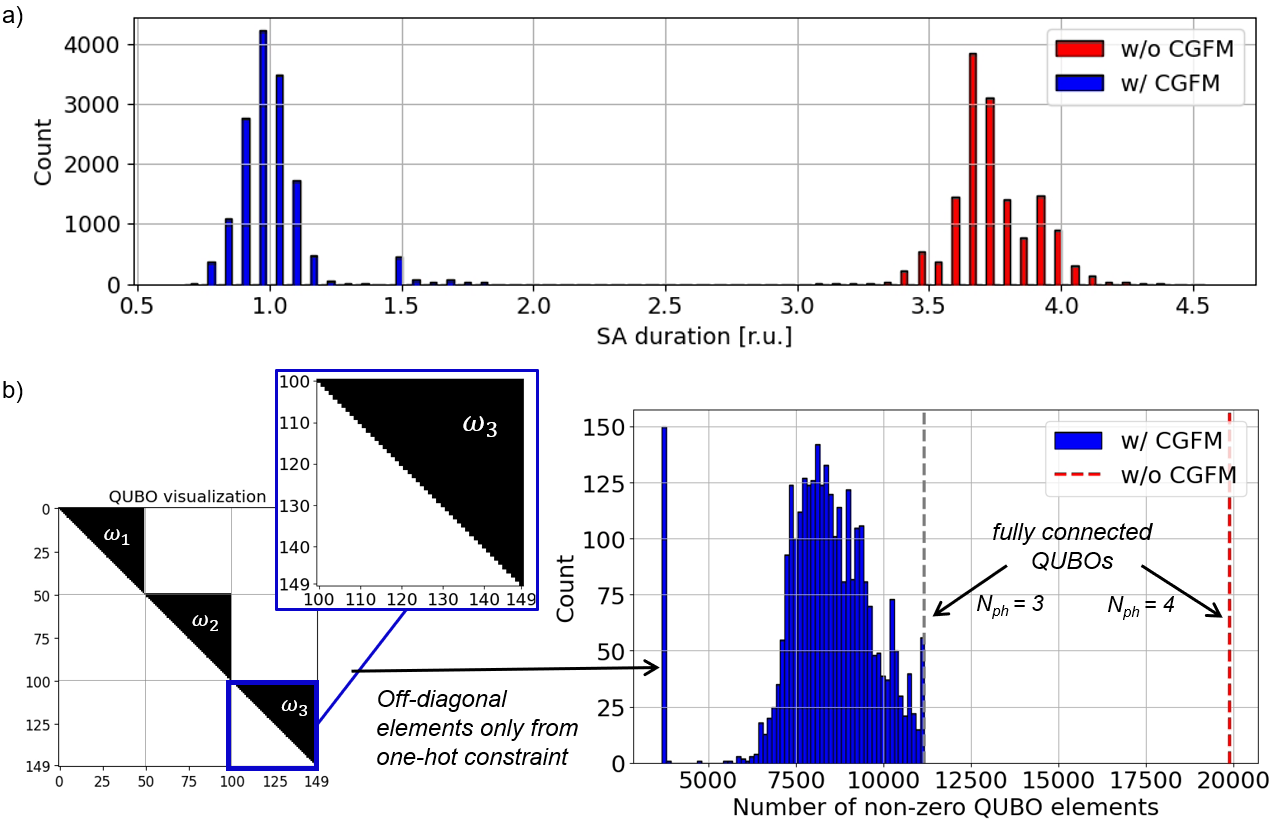}
    \caption{Run time comparisons of the algorithms w/ and w/o CGFM ansatz. a) Histogram 
    comparing the duration of each $5\times5\times600$ single SA optimization runs. The data is collected from each 5 runs of the $\kappa$, $E$, $\rho$, $\delta \alpha$ and $\Delta T$ simulations presented in the main article  Fig.~\ref{fig_results_so_1}. 
    b) Histogram of the number of non-zero QUBO elements in the $5\times5\times600$ QUBOs corresponding to the simulations w/ CGFM ansatz. The reduced number of non-zero elements means less interactions.
    For comparison, the red dashed line indicates the number of non-zero elements of the fully connected QUBO with $N_\text{ph}=4$ (19900) which always results from the approach w/o CGFM owing to the constraint, Eq.~\ref{eq_constraint_encoded}. The gray dashed line marks the maximum number of non-zero elements (11175) in the case of the CGFM ansatz, where $N_\text{ph}=3$. On the left, the QUBO matrix structure of the prominent peak at 3675 non-zero elements is illustrated, where a white (black) square means a zero (non-zero) value. In theses cases, the FM training provides a linear surrogate model and only the added interaction terms related to the one-hot encoding are non-zero. As indicated, each of the $N_\text{ph}=3$ fully interacting bit sets represents one of the 3 discretized constraint-guided variables, $\omega_i$ (e.g., $\omega_3:\small\{x_{100}, x_{101},...,x_{149}\small\}$).}
    \label{fig_speed_up}
\end{figure}

Let us now focus on the costs of the  optimization  using SA. Both approaches, w/ and w/o CGFM, are tested under the same SA parameters, i.e. the number of repetitions (“runs”) and variations (“sweeps”). Each sweep consists of a sequential Metropolis update of all individual binary variables. For the sake of simplicity, we assume that the duration of the update process is the same in the accepting and rejecting case. Owing to the reduced amount of bits, we then would expect an advantage of the method w/ CGFM, which scales again with $N_{\rm ph}/(N_{\rm ph}-1)$ (i.e., a 1.3x speed-up).
Interestingly, when measuring the total wall-clock time of our different full optimization runs (over 600 active learning iterations), we obtain a greater speed-up: The FM+QO method w/ CGFM is 1.6x faster than w/o CGFM, and we find the reason for that in the run time of the isolated QO depending on the inner QUBO structure. Therefore,  Fig.~\ref{fig_speed_up}a illustrates a histogram of the isolated SA algorithm run times, where the data is taken equally from the five scenarios optimizing $\kappa$, $E$, $\rho$, $\delta \alpha$ and $\Delta T$ (see main article Fig.~\ref{fig_results_so_1}a-e). We realize that the CGFM ansatz even provides a speedup of 3.6x for the isolated QO task.
In particular, up to this point, our estimation takes not into account the particular inner QUBO structure. Namely, the SA algorithm internally calculates the cost value of a state in order to reject or accept the Metropolis update event, and that numerical matrix multiplication benefits from sparsity of the QUBO. Based on the algorithm w/o CGFM ansatz always all interactions become non-zero owing to the constraint, Eq.~\ref{eq_constraint_encoded}, i.e., the number of non-zero elements is here always 19900. Instead, when analyzing the trained QUBO models from the runs using CGFM, we find on average only 8351 QUBO interaction terms being non-zero, where a fully connected QUBO would mean 11175 non-zero elements. The distribution of how often the FM modeling results in a QUBO with a certain number of non-zero elements is given in Fig.~\ref{fig_speed_up}b compared against the cases of fully connected QUBOs. Interestingly, we further record about 150 QUBO models which contain only 3675 interaction terms (Fig.~\ref{fig_speed_up}b, isolated blue bin). In these prominent cases, the FM training results in a linear model and only the required interaction terms coming from the one-hot constraint, Eq.~\ref{eq_constr_one_hot}, appear with non-zero values. The QUBO matrix structure corresponding to this special case is visualized on the left-hand side of Fig.~\ref{fig_speed_up}b, where the black squares indicate non-zero elements. This observation seems not correlated to the particular scenario, as we observed such cases in all of our $\kappa$, $E$, $\rho$, $\delta \alpha$ and $\Delta T$ optimizations.

To conclude, utilizing the CGFM ansatz results here in an overall 1.6x speed-up of our FM+QO experiments independent of the optimization scenario. This acceleration arises from two CGFM characteristics, i) search space dimensionality reduction and ii) a clear trend towards sparser QUBO learning models. In classical calculus, a sparser QUBO means less floating point operations. In general, less interaction terms means less local energy minima trapping numerical optimization progress. Therefore, we argue that also other QUBO optimization techniques like QA or QAOA would benefit from the CGFM ansatz, in particular, since qubit connectivity on current hardware is limited.

\section{Brute force search: Density of states on solution space grid}
\label{sec_brute_force_appendix}

\begin{figure}
    \centering
    \includegraphics[width=\textwidth]{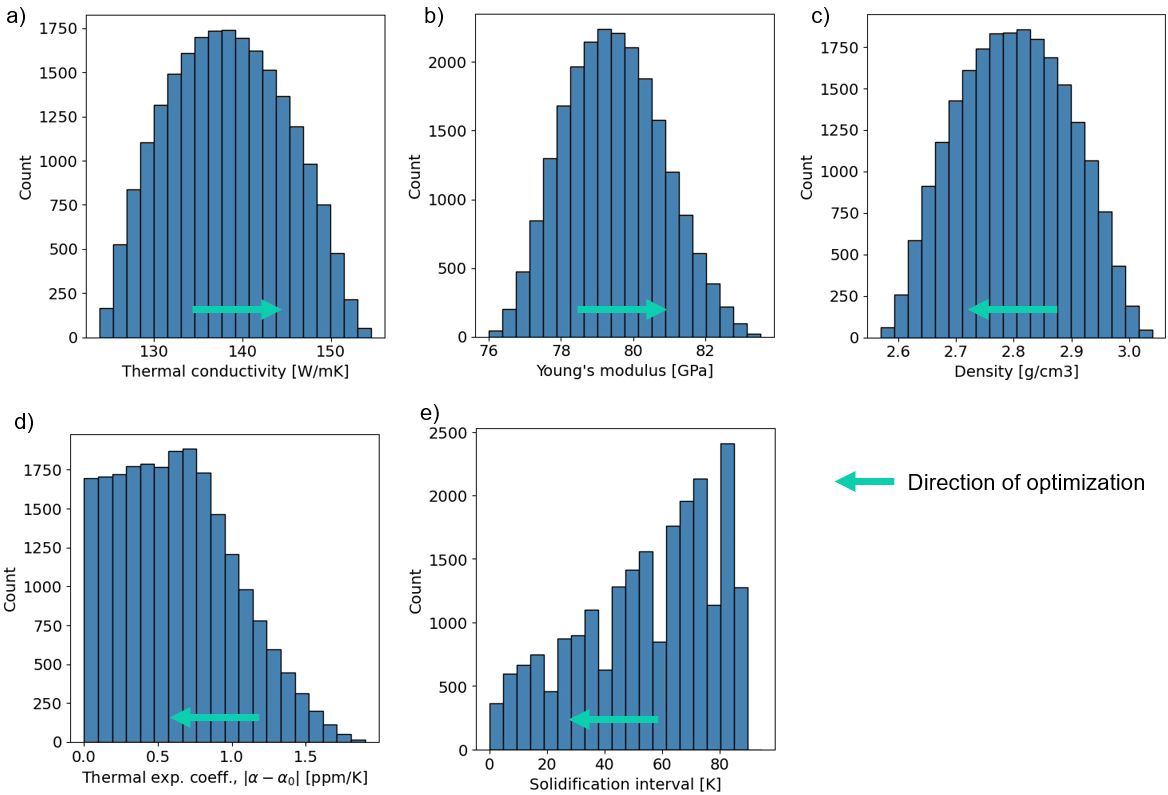}
    \caption{Histogram chart of all alloy designs obtained from brute force search. Panel a-e depict the solution densities depending on thermal conductivity, Young's modulus, mass density, thermal expansion coefficient and the solidification interval, respectively. The cyan colored arrow indicates the optimization direction (i.e., left to right meaning maximization).}
    \label{fig_sol_densities}
\end{figure}

A brute force search is executed in order to examine the density of states (alloy designs) close to the optimum regarding each of our four target properties. Therefore our analytical prediction models are directly employed (introduced in the main article Eqs.~\ref{eq_dalpha}-\ref{eq_deltaT}). However, all alloy designs are taken from a grid of phase fractions, $\bar f_k \in [0,1]$, with 0.02 resolution (i.e., with 50 equal-sized increments). The obtained distributions are illustrated in Fig.~\ref{fig_sol_densities}. The cyan colored arrow indicates the optimization direction. 

Clearly, the solution density is minimal near the optimum in cases of thermal conductivity, Young's modulus and density. This is different for the thermal expansion coefficient and the solidification interval. In particular, the high solution density close to the minimum of the thermal expansion coefficient may suggests a rather shallow global optimum (though we cannot argue about the connectivity of these states from this point of view).

\section{Brute force search: Exact discretized Pareto front}
\label{sec_pareto_brute_force}

\begin{figure}
    \centering
    \includegraphics[width=\textwidth]{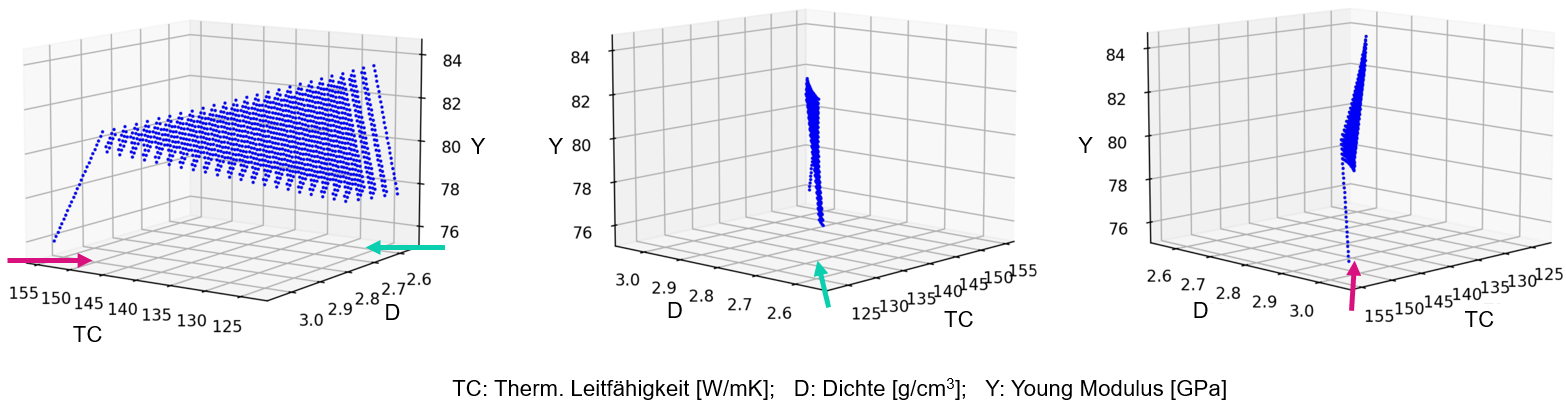}
    \caption{  The exact discretized Pareto fronts corresponding to  Fig.~\ref{fig_results_mo_1} discussed in the main article Sec.~\ref{sec_mo_opt}. It is obtained by brute force search employing the property prediction models (see main article Eqs.~\ref{eq_dalpha}-\ref{eq_deltaT}) and a 0.04 meshing of the phase variables, $\bar f_k$. Different points of view are provided. The colored arrows indicate the point of view with respect to the left figure. As presented the front is not absolutely planar but describes a slightly twisted "sail".}
    \label{fig_results_mo_1_exact}
\end{figure}

In the main article Sec.~\ref{sec_mo_opt} we focus on the multi-objective optimization scenario with the objectives $\kappa$, $E$ and $\rho$, where the Pareto fronts from the FM+QO approaches w/ and w/o DDTS are discussed with Fig.~\ref{fig_results_mo_1}. For comparison, Fig.~\ref{fig_results_mo_1_exact} provides the corresponding exact Pareto front obtained by applying brute force searching as explained in Sec.~\ref{sec_brute_force_appendix} (here based on 0.04 meshing of the phase variables, $\bar f_k$, similar to Sec.~\ref{sec_mo_opt}). Noteworthy, the algorithm works not data-driven and without optimized next candidate search, instead it directly facilitates our analytical property prediction models to scan the entire discretized search space (see Eqs.~\ref{eq_dalpha}-\ref{eq_deltaT} in the main article).

\section{FM+QO: Further Pareto optimization scenarios} 
\label{sec_app_1}

Our Al-alloy model with five material properties allows also for other multi-objective optimization scenarios. 
Therefore, we demonstrate here our DDTS ansatz on three more scenarios against the current state-of-the-art with the weighted-sum scalarization. 

\begin{figure}
    \centering
    \includegraphics[width=\textwidth]{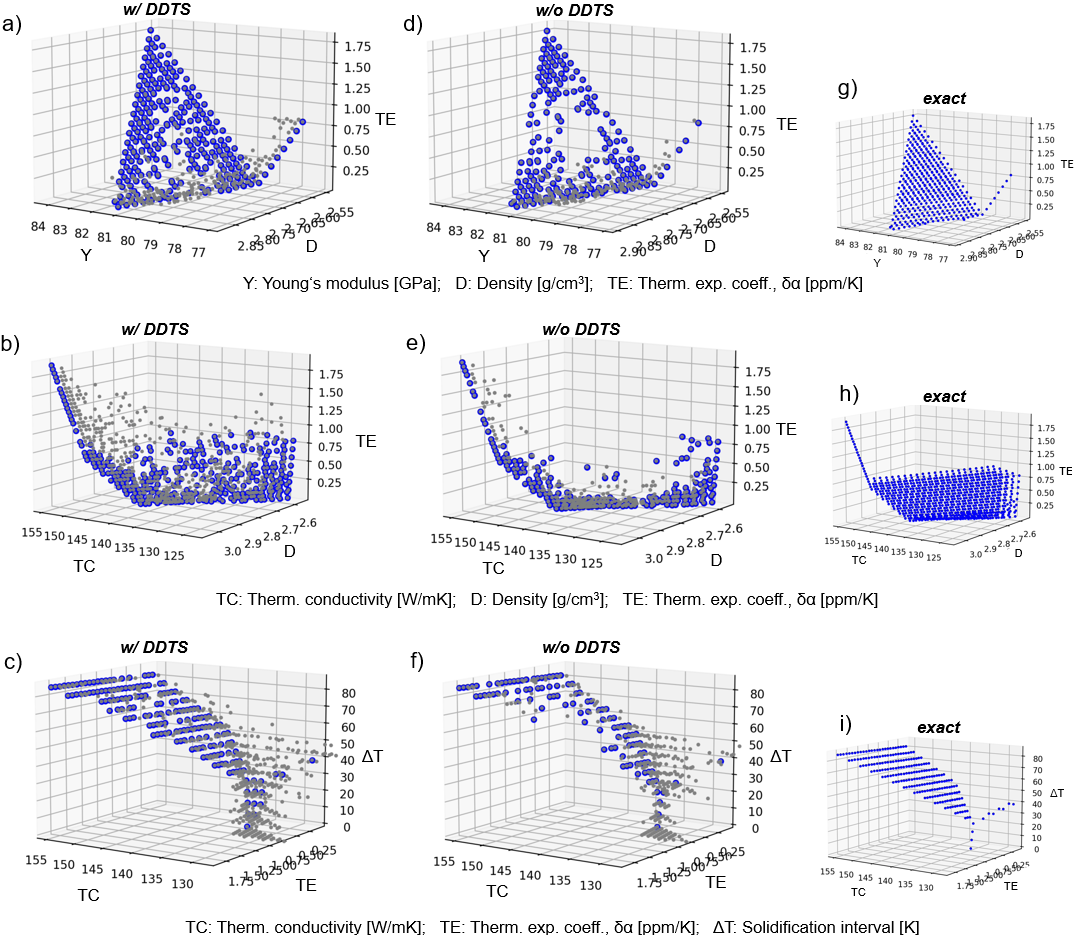}
    \caption{Multi-objective optimization results with the FM+QO method based on either the DDTS ansatz (w/ DDTS) or the standard weighted-sum scalarization (w/o DDTS). The 3d objective space is spanned by three of our five material properties (thermal conductivity, Young's modulus, density, thermal expansion coefficient and solidification interval), each different in pairs (a and d), (b and e), and (c and f). The simulation conditions are similar with runs discussed in the main article Sec.~\ref{sec_mo_opt} (see Fig.~\ref{fig_results_mo_1}). Gray points give the single best alloy designs found by the QO. Blue points indicate the Pareto front (in terms of the set of non-dominating solutions out of the gray points). g-i: Corresponding exact Pareto fronts obtained by brute force search.}
    \label{fig_results_mo_1_appendix}
\end{figure}

We apply similar computational settings to the study discussed in the main article Sec.~\ref{sec_mo_opt}, but interchanged different combinations of the material properties. The resulting optimization solutions (gray dots) and the Pareto fronts (blue dots) are visualized in Fig.~\ref{fig_results_mo_1_appendix}. The results from employing the FM+QO w/ DDTS are given in a-c, while d-f depict analog outcomes from FM+QO with weighted sum scalarization. The respective exact Pareto fronts are shown for comparison in panels g-i.
According to results demonstrated in the main article, applying the DDTS ansatz delivers in all three scenarios again a homogeneous distribution of Pareto optimal alloy designs across the entire front. Contrarily, following ansatz w/o DDTS the found solutions and Pareto optimal designs concentrate on the "peripheral" regions of the front and avoid the interior of the fronts.

\section{\textit{Continuous variable} optimization: Single-objective  optimal alloy designs}
\label{sec_app_best_designs}

Tab.~\ref{table_analy_sol} collects the continuous variable optimal Al-alloy designs for our five material properties: thermal conductivity ($\kappa$, to be maximized, main article Eq.~\ref{eq_k}), Young's modulus ($E$, to be maximized,  main article Eq.~\ref{eq_E}), density ($\rho$, to be minimized, main article Eq.~\ref{eq_rho}), thermal expansion coefficient difference to 20 ppm/K ($\delta\alpha$, to minimized, main article Eq.~\ref{eq_dalpha}) as well as solidification interval ($\Delta T$, to be minimized, main article Eq.~\ref{eq_deltaT}). The unique optimal alloy designs for $\kappa$, $E$ and $\rho$ are straightly found. The optimal solutions regarding  $\delta\alpha$ in continuous space can be determined by translating the minimization problem to a LP. The $\delta\alpha$-optimal alloy design given in Tab.~\ref{table_analy_sol} is only an example solution giving $\delta\alpha=0.0$. As the set of $\delta\alpha$-optimal solutions forms a convex polytope, there exist an infinite number of continuous variable solutions. In particular, $\delta\alpha=0.0$ can also be achieved with minor fractions of Al$_3$Ni: $0 \leq f_\text{Al$_3$Ni}  \leq 0.0019  $, or Al$_2$Cu: $0 \leq  f_\text{Al$_2$Cu}  \leq 0.0013 $ (i.e., with for instance up to  0.19 \% Al$_3$Ni). Furthermore, please note that in the case of minimizing $\Delta T$ only the proper fraction of the eutectic Si phase matters,  $f_\text{Si} = 0.128$. Therefore, the $\Delta T$-optimal alloy design given in Tab.~\ref{table_analy_sol} is also only one of an infinite number of continuous variable solutions.

Most clearly one notices the action of holding the Al-matrix at 80 \% when optimizing $\kappa$, $E$ and $\rho$. In particular, maximizing $\kappa$ and $E$ as well as minimizing $\rho$ runs each towards the interval boundary of a single 20 \% phase fraction. Removing the constraint instead, e.g., by taking also the Al-matrix fraction as a free system variable, the optimal alloy designs would be pure Al (for maximizing $\kappa$), pure Si (for maximizing $E$) and pure Mg$_2$Si (for minimizing $\rho$). This simple designs follow straightly from the simple mixing rules, which we employ here for predicting the properties (see also Tab.~\ref{table_param} in the main article for the phase parameters).

\begin{table}
\centering
\begin{tabular}{ |l||c|c|c|c|c|  }
 \hline
 \multicolumn{6}{|c|}{Analytical single-objective optimal alloy designs} \\
 \hline
  OPT w.r.t  & \textbf{$\kappa$}  & \textbf{$E$}  & \textbf{$\rho$}  & \textbf{$\delta\alpha$}  & \textbf{$\Delta T$} \\
 \hline
 Al (matrix) & 80 \% & 80 \% & 80 \% & 80 \% & 80 \% \\
eutectic Si & 0 \% & 20 \% & 0 \% & 0.0569 \% & 12.8 \%\\
 Mg$_2$Si & 0 \% & 0 \% & 20 \% &  0.1431 \% & 7.2  \%\\
 Al$_3$Ni & 0 \% & 0 \% & 0 \% & 0 \% & 0 \% \\
Al$_2$Cu & 20 \% & 0 \% & 0 \% & 0 \% & 0 \%\\
 \hline
 $\kappa$ [W/m K]   & \textbf{154.5}     & 124.67 & 123.27 &123.67 & 124.17 \\
 $E$ [GPa]          & 75.99     & \textbf{83.67} & 77.60 & 79.33 & 81.48 \\
 $\rho$ [g/cm$^3$]  & 3.04     & 2.62 &  \textbf{2.56} & 2.58 & 2.60 \\
 $\delta\alpha$ [ppm/K] & 1.90 & 1.80 & 0.80 & \textbf{0.0} & 0.93 \\
 $\Delta T$ [K]     & 85.7     & 100.0 & 85.7 & 47.61 & \textbf{0.0} \\
 \hline
\end{tabular}
\caption{Single-objective optimal alloy designs \textit{assuming a continuous variable problem} and regarding the material properties: thermal conductivity ($\kappa$, to be maximized, main article Eq.~\ref{eq_k}), Young's modulus ($E$, to be maximized, main article Eq.~\ref{eq_E}), density ($\rho$, to be minimized, main article Eq.~\ref{eq_rho}), thermal expansion coefficient difference to 20 ppm/K ($\delta\alpha$, to minimized, main article Eq.~\ref{eq_dalpha}) as well as solidification interval ($\Delta T$, to be minimized, main article Eq.~\ref{eq_deltaT}). The upper part of the table declares the optimal design, the lower body of the table the corresponding material properties. The header of the columns indicates the optimization objective. The optimal single-objective values are marked bold. Please note, that the continuous variable problems to minimize $\delta\alpha$ and $\Delta T$ results in an infinite number of solutions with $\delta\alpha=0$ and $\Delta T=0$, respectively. Here, only a single sample design is presented each. Please note further that the optimization is constrained by fixing the fraction of the Al-matrix to 80 \%.}
\label{table_analy_sol}
\end{table}

\newpage

\bibliographystyle{naturemag} 
\bibliography{main}